\documentclass[openacc]{rsproca_new}

\usepackage{graphicx}
\usepackage{amssymb}
\usepackage{amsmath}
\usepackage{psfrag}
\usepackage{mathtools}
\usepackage{color}
\usepackage{xcolor}
\usepackage{lineno}
\usepackage[export]{adjustbox}
\newcommand*\patchAmsMathEnvironmentForLineno[1]{%
  \expandafter\let\csname old#1\expandafter\endcsname\csname #1\endcsname
  \expandafter\let\csname oldend#1\expandafter\endcsname\csname end#1\endcsname
  \renewenvironment{#1}%
     {\linenomath\csname old#1\endcsname}%
     {\csname oldend#1\endcsname\endlinenomath}}%
\newcommand*\patchBothAmsMathEnvironmentsForLineno[1]{%
  \patchAmsMathEnvironmentForLineno{#1}%
  \patchAmsMathEnvironmentForLineno{#1*}}%
\AtBeginDocument{%
\patchBothAmsMathEnvironmentsForLineno{equation}%
\patchBothAmsMathEnvironmentsForLineno{align}%
\patchBothAmsMathEnvironmentsForLineno{flalign}%
\patchBothAmsMathEnvironmentsForLineno{alignat}%
\patchBothAmsMathEnvironmentsForLineno{gather}%
\patchBothAmsMathEnvironmentsForLineno{multline}%
}

\definecolor{mygreen1}{rgb}{0.3, 0.6, 0.1}
\definecolor{lightblue1}{cmyk}{0.47,0,0.13,0}
\definecolor{purple1}{cmyk}{0.5,0.72,0.04,0}
\definecolor{green1}{cmyk}{0.72,0,1,0}
\definecolor{orange1}{cmyk}{0,0.46,0.77,0}
\definecolor{brown1}{cmyk}{0.18,0.4,0.61,0.06}
\definecolor{darkblue1}{cmyk}{0.97,0.84,0,0}
\definecolor{lightblue2}{cmyk}{1,0,0,0}
\definecolor{krgreen}{cmyk}{0.65,0,1,0}
\definecolor{krpurple}{cmyk}{0.41,0.78,0,0}
\definecolor{krcyan}{cmyk}{0.57,0,0.15,0}
\definecolor{krorange}{cmyk}{0,0.56,0.94,0}
\definecolor{krblue}{cmyk}{0.93,0.75,0,0}
\definecolor{krred}{cmyk}{0, 0.95, 0.92,0}
\definecolor{grey}{cmyk}{0, 0, 0,0.7}
\definecolor{cor}{cmyk}{0.07, 0.31, 0.99,0}
\definecolor{cgr}{cmyk}{0.6,0.12,1,0.01}

\newcommand{\tr}[1]{\textcolor{black}{#1}} 
\newcommand{\fc}[1]{\textcolor{black}{#1}} 
\newcommand{\ei}[1]{\textcolor{black}{#1}} 
\newcommand{\dr}[1]{\textcolor{black}{#1}} 

\newcommand{\vi}[1]{\textcolor{black}{#1}} 
\newcommand{\zw}[1]{\textcolor{black}{#1}} 
\newcommand{\gc}[1]{\textcolor{black}{#1}} 
\newcommand{\nc}[1]{\textcolor{black}{#1}} 

\begin{document}

\title{Coupling-Induced Instability in a Ring of Thermoacoustic Oscillators}

\author{
T. Pedergnana and N. Noiray}

\address{ CAPS Laboratory, Department of Mechanical and Process Engineering, ETH Z{\"u}rich, Sonneggstrasse 3,
8092 Z{\"u}rich, Switzerland}

\subject{Acoustics, Fluid Mechanics, Mathematical modelling}


\corres{T. Pedergnana\\
\email{ptiemo@ethz.ch}\\
N. Noiray\\
\email{noirayn@ethz.ch}}

\begin{abstract}
Thermoacoustic instabilities in can-annular combustors of stationary gas turbines lead to unstable Bloch modes which appear as rotating acoustic pressure waves along the turbine annulus. \dr{The multi-scale, multiphysical nature of the full problem makes a detailed analysis challenging. In this work, we derive a low-order, coupled oscillator model of an idealized can-annular combustor.} The unimodal projection of the Helmholtz equation for the can acoustics is combined with the Rayleigh conductivity, which describes the aeroacoustic coupling between neighboring cans. Using a Bloch-wave ansatz, the resulting system is reduced to a single equation for the frequency spectrum. A linear stability analysis is then performed to study the perturbation of the spectrum by the can-to-can interaction. It is observed that the acoustic coupling can suppress or amplify thermoacoustic instabilities, raising the potential for instabilities in nominally stable systems.
\end{abstract}



\maketitle

\section{Introduction} \label{Section 1: Introduction}
\subsection{Thermoacoustic instability in can-annular combustors}
\zw{Thermoacoustic instabilities} are caused by the constructive interaction of unsteady combustion and the acoustics of the chamber. This dynamic phenomenon is highly undesirable because it crucially restricts the operating range of the engine \cite{keller95} and it remains a major challenge to the development of high-performance, low-emission combustion systems, in particular stationary gas turbines \tr{\cite{POINSOT20171}}. For examples of an experimental investigation of a thermoacoustic instability caused by a practical swirling flame, the reader is referred to \cite{Paschereit20001025}. Nonlinear dynamics of free and forced self-excited turbulent premixed flames are studied experimentally in \cite{LI2013947,BALUSAMY20153229}, respectively. \nc{Passive damping of the resulting pulsations can be achieved by perforated liners \cite{Zhao2011725liner} or Helmholtz dampers \cite{Zhao20091672helmholtz}}. \nc{A discussion on the application of Helmholtz dampers in gas turbine combustors was given in \cite{Bellucci2004271}. Feedback control of combustion oscillations was applied in \cite{Evesque20031709,Li2016} and reviewed in \cite{Dowling2005151}.} \dr{Instabilities can occur due to interaction of the flames with low-frequency  eigenmodes of the combustion chamber, with respect to which the flames are compact (see, e.g.,  \cite{crocco1951aspects,keller1985thermally,schuller_poinsot_candel_2020}), or with high-frequency transversal eigenmodes of the chamber, with respect to which the flames are non-compact (see, e.g., \cite{OCONNOR20151,hummel2016theory,BUSCHHAGEN20195181}). We restrict ourselves to the former case in this work. }

\zw{The study of thermoacoustic instabilities dates back to the work of \cite{Rayleigh1878explanation}.} While much research over the last decade in this field has been devoted to understanding the fundamental phenomena associated with thermoacoustic instabilities in \textit{annular} combustors \fc{(see, e.g., \cite{Noiray2013dynamicnature,Ghirardo2013})}, present high-efficiency H-class gas turbines exclusively feature \textit{can-annular} combustor architectures. In this type of system, combustion takes place in a number of cans (typically 12 or 16), without any thermodynamic coupling between the cans. \tr{Yet, the} annular turbine inlet, common to all cans, provides for \fc{aeroacoustic} coupling between adjacent cans. \fc{Acoustic} coupling also occurs through the plenum, affecting especially azimuthal modes. However, the pressure drop across the can burners can, to a certain extent, decouple the plenum acoustics from the can acoustics. Although not always present in gas turbine designs, crossfire tubes between the cans also allow for acoustic coupling. \zw{There exist concepts for future applications to integrate the first vane into the individual cans, which would acoustically decouple the cans at their outlet \cite{Rosic10}. }

\zw{Only} little literature exists on the subject \tr{of thermoacoustic instabilities in can-annular combustors}. Nonetheless, work performed at Siemens \cite{bethke2002thermoacoustic,krebs2005thermoacoustic,Kaufmann08,Farisco17}, General Electric \cite{Bethke19,MOON2020178,MOON2021295} and Ansaldo Energia Switzerland \cite{ghirardo18,ghirardo2020effect} shows that industry itself has started investigating the physics of can-annular combustors.

In their numerical study, \cite{bethke2002thermoacoustic} use the \tr{finite element method (FEM)} and the Helmholtz equation to describe the effect of the can-to-can coupling. Practical aspects of the design of industrial can-annular combustion chambers are discussed in \cite{krebs2005thermoacoustic}. In \cite{Kaufmann08}, mode shapes measured from a single-can test rig are successfully compared to a model of a quarter of an engine with 16 cans in total. Compressible large-eddy simulations (LES) are employed by \cite{Farisco17} to model the coupling between the fluid dynamics and acoustics to study the reflection coefficient and transfer functions between neighboring cans. 

\cite{Bethke19} combine LES with a reduced-order network model to analyze the dynamics of push-push and push-pull modes in a two-can combustor. \ei{The difference between the push-push and push-pull mode are that the former describes acoustic pressure oscillations where all cans are synchronized (in phase), while for the latter there is a phase difference of $\pi$ between neighboring cans.} In the latter reference, the authors test various mitigation strategies for thermoacoustic instabilities, such as fuel split variations, fuel injection location change from nozzle to nozzle within each can and  cross-talk  blockage. They demonstrate experimentally that these measures significantly reduce the acoustic pressure amplitudes generated by the investigated modes. \cite{MOON2020178} analyze experimentally the thermoacoustic dynamics of a four-can system. From the results, the conclusion is made that ``longitudinal-mode instabilities in a can-annular combustion system will preferentially emerge in the form of out-of-phase interactions''. On the same experimental setup, \cite{MOON2021295} study the effect of rotational asymmetry on the thermoacoustics of their can-annular system. We also mention the recently published experimental study of \cite{JEGAL2020}, who investigate the influence of non-identical flame transfer functions (FTFs) in two coupled can combustors on the development of self-excited thermoacoustic oscillations.

\zw{\cite{ghirardo18} present a network model where an assumed impedance boundary condition (BC) is used to quantify the influence of the purely reactive can-to-can communication on the frequency spectrum and on the nature of the modes that appear in the can-annular combustor. The effect of asymmetry of the FTF, i.e. each of the cans exhibits its own FTF, is also explored in their work. The mode shapes in the cans are computed numerically with the FEM from the Helmholtz equation. A subsequent work investigates the influence of noise and nonlinearities in the same model \cite{ghirardo2020effect}. A similar approach is adopted by \cite{YOON2020115774}, who develops a low-order network model of a can-annular combustor with 12 cans. He uses an empirical expression to model the acoustic coupling, and successfully compares mode shapes obtained from his low-order model to FEM simulations of the Helmholtz equation. In \cite{vonSaldern2020analysis}, a FTF is computed from a solver based on the G equation to model the heat release fluctuations of the flame and the Rayleigh conductivity of a compact circular aperture with bias flow is used for the (purely reactive) acoustic can-to-can coupling. The authors study the linear stability of the modeled can-annular combustor. In their follow-up study, they investigate thermoacoustic limit cycles with the same model \cite{vonSaldern2020nonlinear}.} \nc{The dynamics of two coupled thermoacoustic oscillators under asymmetric forcing is investigated in \cite{Sahay2021}.} 

\nc{In two recent studies, the can-annular system is simplified to a network model, where the azimuthal pressure dynamics are represented by the coupling of longitudinal acoustic modes through compact apertures \cite{Fournier2021_1,Fournier2021_2}. In the latter study, the same Rayleigh conductivity is used as in \cite{vonSaldern2020analysis}, and the model equations are simplified using Bloch boundary conditions to study the coupling in more detail. Focusing on reactive coupling effects, the modeled phase response of the connecting gap is successfully compared to experiments.}

\begin{figure}[!t]
\begin{psfrags}
\psfrag{a}{$0$}
\psfrag{b}{$15$}
\psfrag{c}{$30$}
\psfrag{d}{$0$}
\psfrag{e}{$\pi$}
\psfrag{f}{$2\pi$}
\psfrag{g}{$A$}
\psfrag{h}{$A_a$}
\psfrag{i}{$L$}
\psfrag{j}{$d$}
\psfrag{k}{$L$}
\psfrag{l}{
\textbf{A} Helmholtz mode  }
\psfrag{m}{\begin{tabular}{@{}l@{}}
\textbf{B} Half-wave, single cavity ($\lambda\approx 2L$) \\
$\dfrac{D}{L}$ small, nearly closed ($ \mathcal{R}< \dfrac{D}{L}< 1$)
\end{tabular}}
\psfrag{n}{\begin{tabular}{@{}l@{}}
\textbf{C} Half-wave, double cavity ($\lambda\approx 4L$) \\
$\dfrac{D}{L}$ small, nearly open ($\mathcal{R}\approx 1$)
\end{tabular}}
\psfrag{o}{\textbf{A}}
\psfrag{p}{\textbf{B}}
\psfrag{q}{\textbf{C}}
\psfrag{r}{$0.1$}
\psfrag{s}{$0.1$}
\psfrag{t}{$\dfrac{D}{L\mathcal{R}}$}
\psfrag{u}{$5$}
\psfrag{v}{$x$}
\psfrag{w}{$\textcolor{cgr}{\boldsymbol{\tan{\mathrm{H}_L}}}=\textcolor{cor}{\boldsymbol{\dfrac{1+ \sqrt{5}}{\mathrm{H}_L \frac{D}{L\mathcal{R}} }}}$}
\psfrag{A}{$0.2$}
\psfrag{x}{$\mathrm{H}_L$}
\psfrag{B}{$D=d$ $+$ end corrections,\quad $\mathcal{R}=A_a/A$}
\psfrag{C}{$u=0$}
\psfrag{Z}{\begin{tabular}{@{}l@{}}
 Weak \\
coupling
\end{tabular}}
\psfrag{W}{\begin{tabular}{@{}l@{}}
 Strong \\
coupling
\end{tabular}}

\centerline{\includegraphics[width=0.9\textwidth]{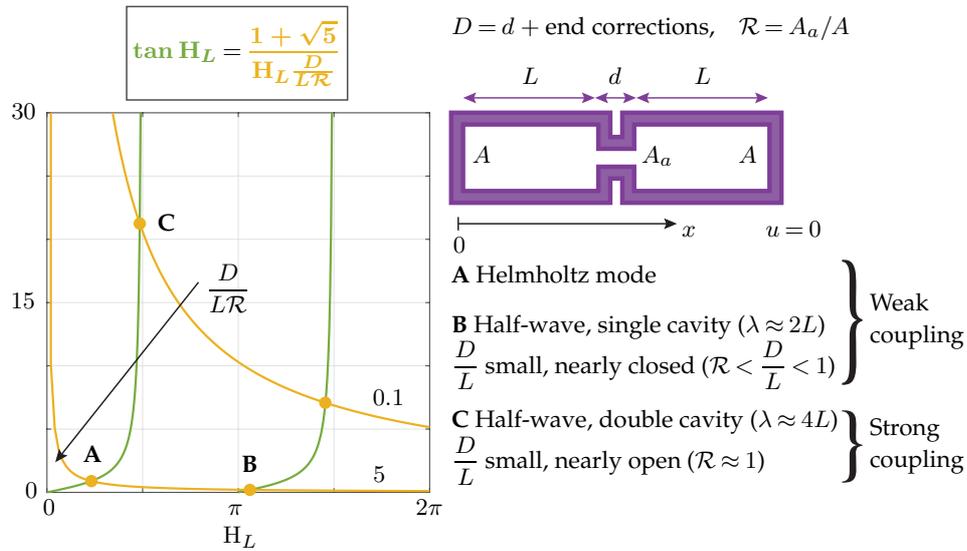}}
\end{psfrags}
\caption{Graphical representation of the equation defining the eigenfrequencies of two coupled cavities of cross-section area $A$ connected by a duct of cross-section area $A_a$. Shown are the RHS and LHS of Eq. \eqref{Compactness condition} for $D /L\mathcal{R}=0.1$ and $5$. Conditions illustrating the three limit cases are denoted by $\textbf{A}$, $\textbf{B}$ and $\textbf{C}$, respectively. $\textbf{A}$ corresponds to the low-frequency coupling involving Helmholtz modes in the cavities. This is a weak coupling scenario. $\textbf{B}$ is also a weakly coupled scenario with half-wavelength modes in the coupled cavities with small aperture (small $\mathcal{R}$) such that $\mathcal{R}<D/L<1$, and $\textbf{C}$ is the case of strong coupling with large aperture and thus $\mathcal{R}$ approaching $1$, which leads to half-wavelength mode along both cavities combined. The present study is concerned only with the limit cases $\textbf{A}$ and $\textbf{B}$.}\label{Figure 0}
\end{figure}

In the present work, a coupled oscillator model is combined with Howe's Rayleigh conductivity of a turbulent wake in a rectangular aperture of thickness $h$ \cite{HOWE1997} to perform a linear stability analysis of an idealized can-annular combustor. \nc{Similar to Ref. \cite{Fournier2021_2}, we assume longitudinal thermoacoustic modes in the cans which communicate through such compact apertures, and we don't resolve in more detail the azimuthal pressure dynamics.} We provide below elementary first-principles calculations to quantify the validity range of this assumption. For this, we consider plane waves propagating in two identical acoustic waveguides of length $L$ closed at one of their ends and connected by a duct, a generic system which is obtained by ``unwrapping'' two coupled cans (see Fig. \ref{Figure 0}). From the linearized mass and momentum balances and with the assumptions of lossless one-dimensional (1D) propagation and compact area expansion at both sides of the connecting duct of length $d$, we can write the transfer matrix between the acoustic pressure and velocity at $x=0$ and at $x=L+d$ as 
\begin{eqnarray}
   && \begin{pmatrix}
p(0)\\
\rho c u(0)
\end{pmatrix}= \nonumber\\
&&\overbrace{\begin{pmatrix}
\cos{kL} \cos{kd}-\mathcal{R} \sin{kL}\sin{kd} & i \cos{kL}\sin{kd}+i\mathcal{R}\sin{kL}\cos{kd}\\
i \sin{kL}\cos{kd}+i\mathcal{R}\cos{kL}\sin{kd} &  -\sin{kL}\sin{kd}+\mathcal{R}\cos{kL}\cos{kd}
\end{pmatrix}}^{\boldsymbol{M}}\begin{pmatrix}
p(L+d)\\
\rho c u(L+d)
\end{pmatrix}, \nonumber
\end{eqnarray}
where $p=p(x)$ and $u=u(x)$ are the acoustic pressure and velocity, respectively, $x$ is the longitudinal coordinate, $k$ is the wavenumber, $\rho$ is the ambient density, $c$ is the ambient speed of sound, $\mathcal{R}=A_a/A$ is the area ratio, $A$ is the cross-section area of the waveguides and $A_a$ is the cross-section area of the connecting duct. Under the same assumptions, we can then write the transfer matrix $\mathcal{M}$ between $x=0$ and $x=2L+d$, where
\begin{eqnarray}
\boldsymbol{\mathcal{M}}=\begin{pmatrix}
M_{11}\cos{kL}+M_{12}(i\sin{kL}/\mathcal{R}) & M_{11}(i\sin{kL})+M_{12}(\cos{kL}/\mathcal{R})\\
M_{21}\cos{kL}+M_{22}(i\sin{kL}/\mathcal{R})&  M_{21}(i\sin{kL})+M_{22}(\cos{kL}/\mathcal{R})
\end{pmatrix}.
\end{eqnarray} 
If now $kd\ll 1$, the following approximation holds:
\begin{equation}
    \boldsymbol{M}\approx \begin{pmatrix}
\cos{kL}-kd \mathcal{R}\sin{kL} & i kd \cos{kL}+i\mathcal{R}\sin{kL}\\
i \sin{kL}+i k d\mathcal{R}\cos{kL} &  -k d \sin{kL}+\mathcal{R}\cos{kL}
\end{pmatrix}.
\end{equation}
For brevity, we write now $p(0)\rightarrow p_1$, $u(0)\rightarrow u_1$, $p(2L+d)\rightarrow p_2$ and $u(2L+d)\rightarrow u_2$. Assuming $u_2=0$ (velocity node at the end of the second can), the normalized impedance $Z_1=p_1/\rho c u_1$ is given by the ratio $\mathcal{M}_{11}/\mathcal{M}_{21}$. If we assume that the upstream end of the first can is also closed, then $Z_1\rightarrow \infty$, or $\mathcal{M}_{21}\rightarrow 0$, which leads to the following characteristic equation:
\begin{equation}
    2\sin{kL}\cos{kL}+kd\Big[\mathcal{R}\cos^2{kL}-\dfrac{1}{\mathcal{R}}\sin^2{kL}\Big]=0, \label{Compactness condition}
\end{equation} 
whose roots are the eigenfrequencies of the pair of coupled cans sketched in Fig. \ref{Figure 0}. We now examine in which parameter range this condition will be satisfied.

We begin with the limit case of completely separated cavities, $\mathcal{R}\rightarrow 0$. One solution satisfying Eq. \eqref{Compactness condition} is $\sin{kL}=0$, which corresponds to the half-wavelength mode of a single can, with wavelength $\lambda\approx2L$. If now, in addition to $kd\ll 1$, $kL \ll 1$, we have $k=\sqrt{2 A_a/V_\mathrm{w} d}$, where $V_\mathrm{w}$ is the volume of a single waveguide, corresponding to a Helmholtz mode of two resonators in series, each with volume $V_\mathrm{w}$, neck length $d$ and cross section $A_a$.

In the general case, after some algebra and replacing the duct length $d$ with $D=d+2l_c$, where $l_c$ is the end correction at one end of the compact coupling duct of length $d$ \cite{howe2014acoustics}, Eq. \eqref{Compactness condition} can be rearranged to yield $\tan{\mathrm{H}_L}=(1\pm \sqrt{5})/(\mathrm{H}_L \dfrac{D}{L \mathcal{R}} )$, where only the positive root is physically relevant, $\mathrm{H}_L=\omega L/c$ is the Helmholtz number and $\omega=k c$. The end correction can be roughly approximated by the hydraulic radius: $l_c\approx \sqrt{A_a/\pi}$, and in the limit case of a thin orifice $D\rightarrow 2\sqrt{A_a/\pi}$. We note, however, that $l_c$ is also affected by the aeroacoustic coupling between the cans, which may increase or decrease the effective attached mass at the aperture and which will be investigated in the following sections of this paper. 

We now consider a third limit case: For small $D /L \mathcal{R}\ll 1$ but moderate $\mathcal{R}\approx 1$, i.e., when the duct between the waveguides is very short and nearly open, we infer from Eq. \eqref{Compactness condition} that $\mathrm{H}_L\approx \pi/2$, which corresponds to the half-wavelength mode of a double cavity with wavelength $\lambda\approx4L$.

These results are illustrated in Fig. \eqref{Figure 0}. Shown are the right-hand-side (RHS) and left-hand-side (LHS) of Eq. \eqref{Compactness condition} for $D /L \mathcal{R}=0.1$ and $5$. Conditions representative of the limiting cases analyzed above are denoted by $\textbf{A}$, $\textbf{B}$ and $\textbf{C}$, respectively. $\textbf{A}$ corresponds to the weakly coupled case with Helmholtz modes in the cavities, $\textbf{B}$ is the weakly coupled case with half-wavelength modes in the coupled cavities, and $\textbf{C}$ is the case of a nearly open duct, also with $D/L\ll 1$, filled by a half-wavelength mode spanning both cavities. The present study is concerned only with cases $\textbf{A}$ and $\textbf{B}$, which correspond to the weak coupling scenarios, for which the diameter and the effective length of the connecting aperture are small with respect to both the can length and the can diameter. Importantly, we note that $\textbf{A}$, $\textbf{B}$ and $\textbf{C}$ may be identified as push-pull modes, when, in the former two cases, the phase difference between the oscillation of the modes in both cavities happens to be exactly $\pi$, and because in the latter case the acoustic pressure on either side of the aperture always satisfies this phase condition. Therefore, when we denote some phase pattern by the push-pull mode in our study below, this should not be understood in an exclusive sense, but in the context of the parameter range considered in this work ($\mathcal{R}<D/L< 1$).

By taking into account the acoustic-hydrodynamic interaction in the apertures between the cans, we observe resistive effects due to the coupling which have either been neglected or not investigated in detail in previous studies \cite{ghirardo18,ghirardo2020effect,YOON2020115774,vonSaldern2020analysis,vonSaldern2020nonlinear,Fournier2021_1,Fournier2021_2}. Our model describes simultaneously the can acoustics and the turbulent wake dynamics in the apertures. Previous works on thermoacoustic instabilities in can-annular combustors paid less attention to the fluid dynamics underlying the coupling and more to the flame dynamics in the individual cans, which were modeled with more complex FTFs. By shifting the focus to the coupling, we aim to better understand the intriguing thermoacoustics that result from the collective behavior of the cans.

\subsection{Acoustic coupling between neighboring cans}
\vi{For low Mach numbers, the fluid motion in the apertures between the cans can be approximated as incompressible (see p. 33 in \cite{howe_1998}). By Howe's energy corollary, vorticity fluctuations in an incompressible, turbulent shear layer or wake can increase or decrease the acoustic energy of a sound field they interact with \cite{Howe1980407}. In our model, the Rayleigh conductivity $K_R$ describes the interaction between the can acoustics and the turbulent wake in the aperture between the cans. $K_R$ has dimension length. In Howe's theory of flow-excited deep cavity oscillations, positive imaginary and real parts of $K_R$ are associated with amplification of the sound field by the mean flow and reduction of the oscillation frequency, respectively (see p. 443 in \cite{howe_1998}).} 

\dr{To compute $K_R$, we follow Howe's derivation for uniform, two-sided grazing turbulent flow over a rectangular aperture of thickness $h$ \cite{HOWE1997}. Rayleigh conductivity models for different geometries are found, e.g., in chapters 5 and 6 of \cite{howe_1998}. The centerpiece of the model is a turbulent wake of thickness $h$, the can spacing, which is bounded by two vortex sheets. The wake separates the hot gas streams of adjacent cans. This is an idealized model for the turbulent fluid motion in the aperture. The (complex) displacement of the vortex sheet $\zeta$ represents the coherent (at the acoustic frequency) vorticity fluctuations in the aperture that arise from the forced motion of the turbulent wake.} \zw{From the Rayleigh conductivity, we can compute the acoustic impedance, which appears in the BCs of the Helmholtz equation governing the can acoustics \cite{jing1999experimental,SUN2002557,Tiemo20}. }

\begin{figure}[!t]
\begin{psfrags}
\psfrag{a}{\footnotesize  \textcolor{green1}{\bf First row of turbine vanes}}
\psfrag{b}{\footnotesize \textcolor{violet}{\bf Casing of neighbouring cans}}
\psfrag{A}{\hspace{1.5cm} \small \bf  (a) Axial view of turbine inlet}
\psfrag{B}{\hspace{1cm} \small \bf (b) Radial view of can outlet and turbine inlet}
\psfrag{c}{$\tilde{z}$}
\psfrag{d}{$\tilde{y}$}
\psfrag{e}{$\tilde{x}$}
\psfrag{f}{\textcolor{purple1}{$C$}}
\psfrag{g}{\textcolor{purple1}{$B$}}
\psfrag{h}{\textcolor{purple1}{ $h$}}
\psfrag{i}{$z$}
\psfrag{j}{$y$}
\psfrag{y}{$x$}
\psfrag{k}{\footnotesize \bf \textcolor{brown1}{ \begin{tabular}{@{}c@{}}
High Mach \\
region
\end{tabular}}}
\psfrag{l}{\hspace{-5.3cm} \bf \footnotesize \textcolor{darkblue1}{\begin{tabular}{@{}c@{}}
Mean \\
axial\\
velocity
\end{tabular}}}
\psfrag{m}{\footnotesize \bf \textcolor{lightblue2}{\begin{tabular}{@{}c@{}}
Turbulent \\
wake
\end{tabular}}}
\psfrag{n}{\footnotesize \bf \textcolor{orange1}{Vortex sheet}}
\psfrag{o}{\textcolor{darkblue1}{$U_\mathrm{tot}$}}
\psfrag{p}{\textcolor{darkblue1}{$0$}}
\psfrag{q}{\hspace{0.1cm}\footnotesize \bf \textcolor{gray}{ \begin{tabular}{@{}c@{}}
Low Mach \\
region
\end{tabular}}}
\psfrag{r}{ \textcolor{orange1}{ $\mathrm{Re}(\zeta)$}}
\psfrag{s}{$x$}
\psfrag{t}{\textcolor{purple1}{$W$}}
\psfrag{u}{$y$}
\psfrag{v}{\textcolor{purple1}{$h$}}
\psfrag{w}{$z$}
\psfrag{x}{$0$}
\centerline{\includegraphics[width=1\textwidth]{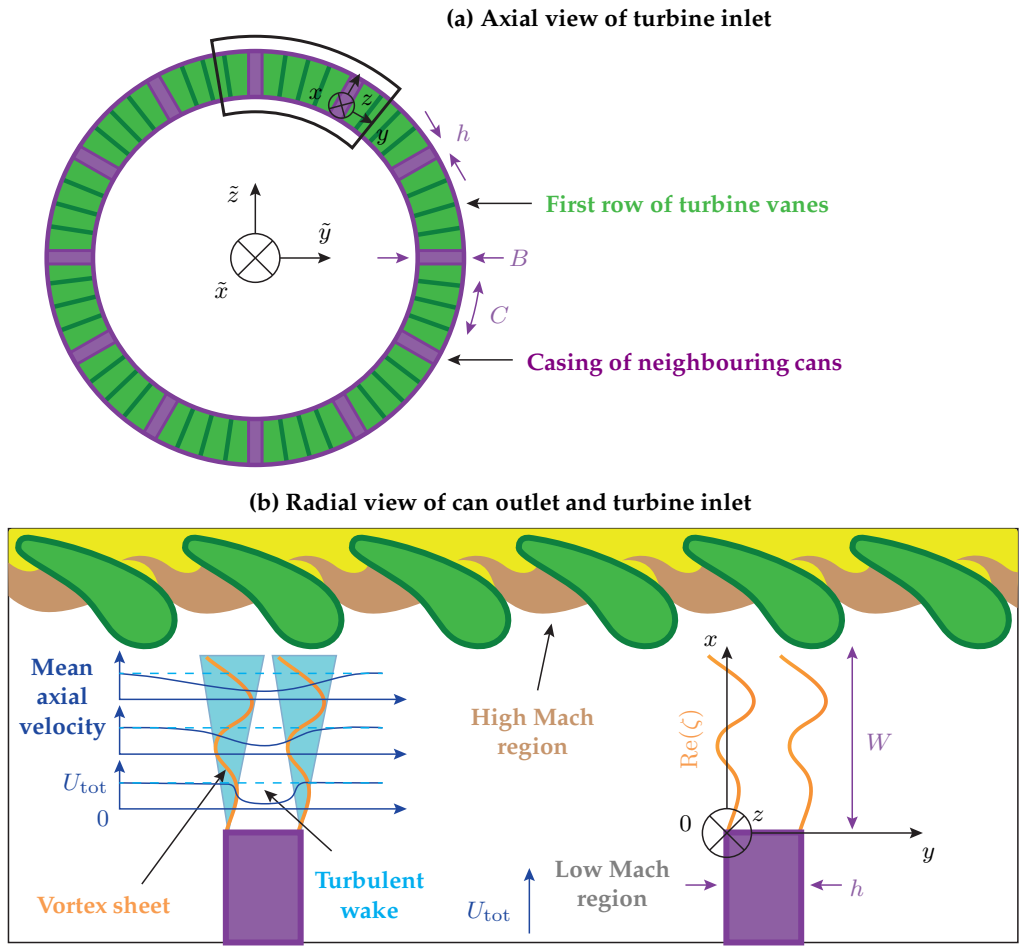}}
\end{psfrags}
\caption{\zw{Sketch of the model for the acoustic coupling between the cans. \textbf{(a)}
Turbine inlet at $\tilde{x}=0$ and the casings of the 12 cans. Shown are the local and global coordinate systems $(x,y,z)$ and $(\tilde{x},\tilde{y},\tilde{z})$, respectively, the thickness $h$ of the casing of neighbouring cans, the height of the coupling aperture $B$ and the width of the cans at the turbine inlet $C$. \textbf{(b)} Sketch of the turbine inlet region, showing a typical mean axial velocity profile in the aperture, the bulk velocity of the combustion products $U_\mathrm{tot}$ and the real part of the vortex sheet displacement $\mathrm{Re}(\zeta)$. }}\label{Figure 1}
\end{figure}

\zw{The model for the acoustic coupling is sketched in Fig. \ref{Figure 1}. Figure \ref{Figure 1}\textbf{(a)} shows the turbine inlet at $\tilde{x}=0$, the casings of the 12 cans, the local and global coordinate systems $(x,y,z)$ and $(\tilde{x},\tilde{y},\tilde{z})$, respectively, the thickness $h$ of the casing of neighbouring cans, the height of the coupling aperture $B$ and the width of the cans at the turbine inlet $C$. Figure \ref{Figure 1}\textbf{(b)} shows a typical mean axial velocity profile in the aperture, the bulk velocity of the combustion products $U_\mathrm{tot}$ and the real part of the vortex sheet displacement $\mathrm{Re}(\zeta)$.}

\zw{In the model, vorticity disturbances are advected at a constant mean axial velocity $U<U_\mathrm{tot}$. We assume that $U$ and $U_\mathrm{tot}$ are related by $U=U_\mathrm{tot}/2$. This approximation accounts for the sharp drop-off of the mean flow speed in the presence of the turbulent boundary layer at the wall.} \ei{Due to this drop-off, vorticity fluctuations in the aperture are advected at a lower speed than the bulk velocity $U_\mathrm{tot}$.} 

The assumption that $U=U_\mathrm{tot}/2$ is now briefly discussed. According to Howe, ``the fluid within the volume of the aperture [$\dots$] is  assumed  to  be  in  a  mean  state  of  rest'' \cite{HOWE1997}. This motivates the choice for $U$: inside the aperture, $U\approx0$, while far away from the wall, $U\approx U_\mathrm{tot}$. At the edge of the turbulent wake, we approximate $U$ by $U_\mathrm{tot}/2$, which is in agreement with classical estimates in literature (see p. 456 in \cite{howe_1998}). To further justify this assumption, we compare Fig. 10 in \cite{CHEN2020115547} (measured acoustic impedance of a rectangular slot) to Fig. 4 in \cite{Howe1996} (Rayleigh conductivity for a rectangular slot). In the former work, $\mathrm{Re}(Z)$ has a first local minimum at around (a) $\omega W/U_\mathrm{tot}\approx2$. In the latter work, $\mathrm{Im}(K_R)$ has a local maximum around (b) $\omega W/2U\approx2.4$. Assuming (a) and (b) describe the same point ($\mathrm{Im}(K_R)>0$ or $\mathrm{Re}(Z)<0$ both imply amplification of the sound field by the mean flow in the aperture \cite{Tiemo20}), this implies $U/U_\mathrm{tot}\approx0.42$ for the experiments of \cite{CHEN2020115547}. We also mention the study in \cite{Tiemo20}, where a Rayleigh conductivity model was calibrated to experimental results to obtain a predictive model of the acoustic impedance of a side branch aperture. After calibration, the value of $U/U_\mathrm{tot}$ obtained therein is within $1$ percent of $0.5$ ($U$ corresponds to $U_-$ in their notation).

\vi{The present study focuses on the perturbation of the frequency spectrum of a ring of thermoacoustic oscillators by mean flow effects on the acoustic coupling. We note that in reality, when a thermoacoustic instability occurs and the sound field reaches finite amplitudes, the acoustics lead to changes of the mean flow itself, and the problem becomes nonlinear. The nonlinear saturation of a forced shear layer over a T-junction by large-amplitude acoustic forcing was studied numerically in \cite{boujo_bauerheim_noiray_2018}. Their results are consistent with aeroacoustic experiments on a similar configuration presented in \cite{bourquard_faure-beaulieu_noiray_2021}. Such nonlinear effects are neglected in the present work, which is restricted to linearized dynamics. \nc{We mention that aeroacoustic characterization of T-junctions based on LES and system identification is performed in \cite{Foller2010}.}}

\subsection{Bloch modes}
\zw{We now turn to a different topic, which is also central to this work. Bloch wave theory was developed in the $20$th century to model the wave-like behavior of electrons in crystal lattices, where they are subject to a periodic potential due to the atoms  \cite{bloch1929quantenmechanik}. A more modern account of the theory is found, e.g., in \cite{kittel1996introduction}. The main result states that in a perfectly periodic Hermitian system, in our case the Helmholtz equation, the wave function, in our case the acoustic pressure $\hat{p}$, can be written as a plane wave with periodic amplitude. }

\cite{mensah2016efficient} have extended Bloch wave theory to thermoacoustic systems by making use of inherent discrete rotational symmetries of typical combustion chamber designs to compute thermoacoustic eigenmodes of an annular combustor. By imposing periodic BCs, they obtain a significant reduction of the computational effort for determining the thermoacoustic modes in their system. Their results were used by \cite{ghirardo18} to derive equivalent BCs in their study of the thermoacoustic modes \zw{in} a can-annular combustor. The same theory was also applied by \cite{Haeringer19} in the time domain to reduce the computational cost of fluid dynamics simulations for the modelling of limit cycle oscillations in (can-)annular combustors. \cite{haeringer2020strategy} employ Bloch wave theory to derive equivalent reflection coefficients that represent the can-to-can coupling. They propose a strategy to impose such reflection coefficients at the acoustic terminations of a single-can test rig by installing passive acoustic elements such as straight ducts or Helmholtz resonators, to mimic the thermoacoustic behavior of a full engine. \zw{In our application of Bloch wave theory, we follow the approach presented by \cite{vonSaldern2020analysis}, who use a Bloch wave ansatz to derive equivalent boundary conditions for a modeled can-annular combustor in the frequency domain. This enables the analysis of a can-annular system by considering a single can, thus reducing the number of equations by a factor $N$. Indirect experimental evidence of Bloch modes in real-world gas turbines is provided by the spectrograms shown in Fig. 8 of \cite{ghirardo18}, where pressure signals from different cans were decomposed into Bloch modes using the discrete Fourier transform. Direct evidence of Bloch modes occurring in a four-can system, showing wave-like phase patterns along the annulus, is presented in Figs. 5 and 6 in \cite{MOON2020178}.}

\zw{Based on the results of \cite{ghirardo18} and \cite{mensah2016efficient}, we identify azimuthal phase patterns in our model, which we call Bloch modes. Different Bloch modes are distinguished by the Bloch wavenumber $b$ which determines the relative phase between the acoustic pressure fields of adjacent cans.}

\dr{When the acoustic pressure is visualized at a fixed axial position, Bloch modes appear as rotating waves spinning around the turbine annulus \cite{emenheiser2016patterns}. These apparent waves can occur because neighboring cans communicate through the apertures at the turbine inlet. The Bloch modes we consider are not ``true'' azimuthal waves, which may arise in the annular plenum before the burner or at the turbine inlet, because the medium is not modeled as a continuum, but as discrete control volumes with individual, but coupled internal dynamics. } 

\ei{Instabilities of azimuthal waves in a discrete fluid-dynamical system are encountered in a different context by \cite{couchman_turton_bush_2019} and \cite{couchman_bush_2020}, who combine theoretical and experimental methods to study of the dynamics of a ring of bouncing droplets.}

\subsection{Overview}

\zw{The paper is structured as follows: We discuss the key assumptions of our study in $\S$\ref{Section: setting}. In $\S$\ref{Section 2: Thermoacoustics}, a coupled oscillator model of an idealized can-annular combustor is derived from a unimodal projection of the Helmholtz equation. Using a Bloch wave ansatz, the resulting system of $N$ ordinary differential equations (ODEs) is then reduced to a single equation for the frequency spectrum. In $\S$\ref{Section 5: Results}, a parameter study is performed on the spectrum to determine the linear stability of the system at different conditions. We discuss and give a physical interpretation of our results. Our conclusions are summarized in $\S$\ref{Section 6: Conclusions}.}

\section{Set-up \label{Section: setting}}

\dr{In the present work, the internal dynamics of the individual cans are simplified to a minimum, and special emphasis is placed on modeling the can-to-can communication. Following \cite{YOON2020115774}, we assume a closed BC at the turbine inlet, where the high Mach number in the first row of turbine vanes leads to full reflection of incident acoustic energy. Interested readers can refer to \cite{WEILENMANN2021115799} for a recent experimental study on sound reflection by high-Mach and choked nozzle flows. A generalized model for nozzles with losses is presented in \cite{DeDomenico2019212}. }

\dr{The cans are modeled as unimodal thermoacoustic oscillators. A linear relation between the acoustic pressure and the coherent heat release rate fluctuations is used, which is justified at small enough acoustic pressure amplitudes \cite{Noiray16}.} \vi{In our model, the flame drives a single natural (longitudinal) eigenmode $\psi_k$ of the can. In a first approximation, we assume that the mode shape of $\psi_k$ is unperturbed by the thermo- and aeroacoustic interactions and that the acoustic pressure signal is close to harmonic. These are often reasonable assumptions in practice \cite{Lieuwen03,culick2006unsteady}.} 

\dr{The above assumptions imply that we restrict ourselves to conditions near the stability limit of a thermoacoustic system where the observed power spectral density (PSD) of the acoustic pressure signal shows an isolated peak near $\omega_k$, which is much larger than all other observed peaks. We note that there can also arise situations where multiple modes are closely spaced, leading to nonlinear interactions between them \cite{Acharya2018309}. We further restrict our study to small perturbations of the frequency spectrum by the coupling, assuming that the thermoacoustic interaction of the sound field and the flames is the dominant source of acoustic energy.}

\dr{Low-order oscillator models of thermoacoustic instabilities have been validated in prior studies performed by our group \cite{Noiray16,boujo2016quantifying,NOIRAY_DENISOV17,BONCIOLINI20195351}, and are well understood in terms of their accuracy compared to higher-order models that include several eigenmodes, explicit time delay and non-antisymmetric nonlinear flame response to acoustic perturbations \cite{BONCIOLINI2021396}. As is shown in Fig. 17 of the latter reference, the simplest model, which is based on a single eigenmode, does not reproduce the PSD over a broad frequency range, but it is sufficient to qualitatively represent the spectral content in the vicinity of the governing eigenfrequency.}

\ei{We approximate the thermoacoustic dynamics in each individual can by the 1D Helmholtz equation with fluctuating heat release rate. Under this assumption, each can exhibits the same internal dynamics as a Rijke tube, albeit with different boundary conditions. Following \cite{ghirardo18}, we neglect low-Mach effects and assume zero mean flow in the can volume. A derivation of the wave equation for this classic example can be found, e.g., in \cite{Maling63}. For a discussion of thermoacoustic oscillations in a Rijke tube in the presence of a temperature gradient, the reader is referred to \cite{rott_1984}. We also mention the more recent studies of self-sustained oscillations in Rijke tubes in Refs. \cite{Matveev2003689,Balasubramanian2008,Juniper2011272,Magri2013183,Rigas2015}.} 

\dr{Let us now briefly discuss some of the simplifying assumptions of our model. First, we neglect the effect of the mean flow on the internal acoustics of the cans but take into its effects on the aeroacoustic coupling between the cans. This approximation, which greatly simplifies our analysis, is in line with our intent to focus on the effect of the aeroacoustic  coupling in the apertures on the linear stability of can-annular combustors.} 

\vi{Secondly, we consider an idealized can-annular combustor where the cans are represented by acoustic waveguides with constant cross-section connected by compact rectangular apertures. This is an abstraction of the typical geometry of a can in an industrial can-annular combustor, which is shown in Fig. 1 of \cite{ghirardo18}. As discussed therein, the cans' cross-section shape changes from circular to nearly rectangular at the turbine inlet while the cross-section area remains roughly constant, and neighboring cans are connected by rectangular apertures whose dimensions are much smaller than the can length.}

\dr{Thirdly, the geometry considered in \cite{HOWE1997} and used in this work is an aperture in an infinite plate of thickness $h$. This idealized configuration does not feature the accelerated flow downstream of the aperture, which is found in the first row of turbine vanes of the present configuration, and it just requires the simple Green's function for infinite half-spaces. We nonetheless take this model as a first approximation of the Rayleigh conductivity in the apertures because (a) the dynamics of the vortex sheets is mainly governed by the Kutta condition at the upstream edge of the apertures \cite{howe_1981} and is weakly influenced by the downstream flow, and (b) the scope of this work is to provide a simplified analysis of the physical phenomena that define the linear stability of can-annular combustors.}

\dr{In future studies, the present thermoacoustic model could be extended to include mean flow effects on the can acoustics and a more detailed representation of the problem geometry. A method for including mean flow effects on zero-Mach thermoacoustic network models is presented in \cite{Motheau2014246}. The Rayleigh conductivity we use could also be refined by extending the model to finite aperture sizes, using more complex Green's functions which take into account finite Mach number effects. This is done in \cite{Yang2016294}, where a semi-analytical model for the acoustic impedance of finite-length circular holes with bias flow is derived by extending the classic model of Howe for acoustically compact holes \cite{howe1979theory}.}

\nc{The present work focuses on the linear stability of the system and we therefore do not investigate nonlinear phenomena pertaining to such can-annular configurations, such as amplitude death and quenching (see, e.g., \cite{Biwa2015,Biwa2016,Thomas18AmpDeathinCoup,Dange2019,Hyodo2020}).}

\section{Model derivation} \label{Section 2: Thermoacoustics}
\subsection{Dynamics of the thermoacoustic system}
In this section, we derive a thermoacoustic model of an idealized can-annular combustor. The system consists of $N$ identical cans, numbered by the integer $j=1,...,N$. We follow the convention that a positive \zw{increment} in $j$ implies a clockwise shift around the streamwise axis. In the following, $\hat{f}$ denotes the Laplace transform \cite{debnath2014integral} of a function $f(t)$, $t\in \mathbb{R}$. To avoid confusion, we use bracketed subscripts on variables to refer to different cans, so that $a_{(j)}$ denotes a variable quantity $a$ in the $j^\mathrm{th}$ can. 

\begin{figure}[!t]
\begin{psfrags}
\psfrag{a}{\textbf{(a)}}
\psfrag{b}{\textbf{(b)}}
\psfrag{c}{\hspace{0.1cm}\textcolor{purple1}{$W$}}
\psfrag{d}{$\tilde{x}$}
\psfrag{e}{\textcolor{purple1}{$L$}}
\psfrag{f}{$0$}
\psfrag{g}{\hspace{0.05cm}\textcolor{red}{\footnotesize \begin{tabular}{@{}c@{}}
Cross-section \\
area $A$
\end{tabular}}}
\psfrag{h}{\footnotesize Burner outlet}
\psfrag{i}{\footnotesize \bf Flame}
\psfrag{j}{\footnotesize Can outlet}
\psfrag{k}{\footnotesize \begin{tabular}{@{}c@{}}
First row of \\
turbine vanes
\end{tabular}}
\psfrag{l}{$\hat{p}_{(j)}$}
\psfrag{m}{$\hat{u}_{(j)}$}
\psfrag{n}{$\sigma_j$}
\psfrag{o}{$\mathcal{V}_{j}$}
\psfrag{p}{$\hat{Q}$}
\psfrag{q}{$\hat{p}_{d,(j-1)}$}
\psfrag{r}{\textcolor{orange}{$\sigma_{2,j}$}}
\psfrag{s}{\textcolor{orange}{$\hat{u}_{(j-1,j)}$}}
\psfrag{t}{\textcolor{cyan}{$\sigma_{4,j}$}}
\psfrag{u}{\textcolor{red}{$\sigma_{1,j}$}}
\psfrag{v}{\textcolor{red}{$\hat{u}_{d,(j)}$}}
\psfrag{w}{$\hat{p}_{d,(j)}$}
\psfrag{x}{\textcolor{orange1}{$\sigma_{3,j}$}}
\psfrag{y}{\textcolor{orange1}{$\hat{u}_{(j,j+1)}$}}
\psfrag{z}{$\hat{p}_{d,(j+1)}$}
\centerline{\includegraphics[width=1\textwidth]{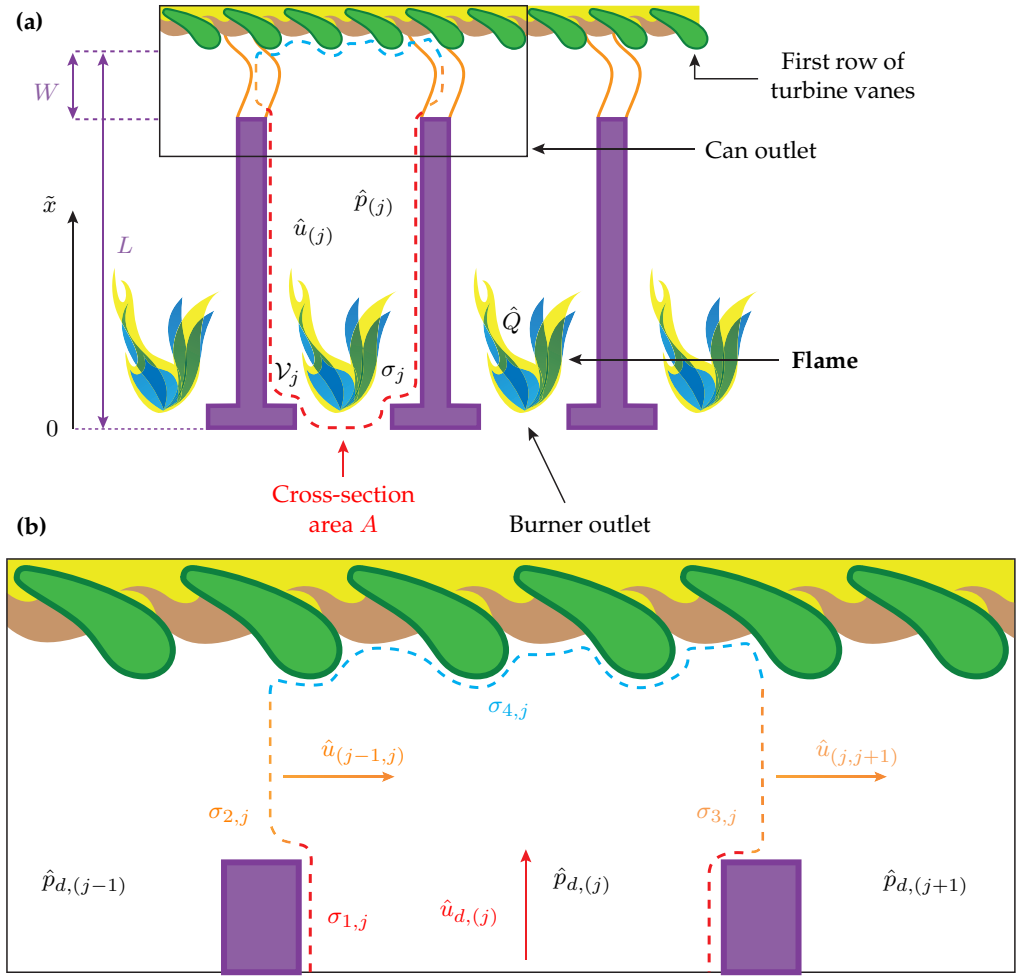}}
\end{psfrags}
\caption{\zw{Sketch of the thermoacoustic model of the $j^{\mathrm{th}}$ can. The dimensions are not true to scale. In \textbf{(a)}, $\mathcal{V}_{j}$ denotes the control volume, $A$ the cross-section area and $L$ the length of the can, respectively. $\hat{Q}$ is the unsteady heat release rate across the flame front. The boundary $\sigma_j$ is divided into 4 parts: $\sigma_{1,j}$ at the can walls and burner outlet, $\sigma_{2,j}$ and $\sigma_{3,j}$ at the coupling interfaces, where the $j^{\mathrm{th}}$ can is connected to the $(j-1)^{\mathrm{th}}$ and $(j+1)^{\mathrm{th}}$ cans, respectively, and $\sigma_{4,j}$ at the turbine inlet at $\tilde{x}=L$. The acoustic velocity in streamwise direction and the acoustic pressure in the $j^{\mathrm{th}}$ can are denoted by $\hat{u}_{(j)}$ and $\hat{p}_{(j)}$, respectively, and $\hat{u}_{d,(j)}$ and $\hat{p}_{d,(j)}$ denote these quantities in the downstream section of each can, i.e., immediately upstream of the outlet of the cans. In \textbf{(b)}, the transverse acoustic velocity on $\sigma_{2,j}$, which results from the pressure difference $\hat{p}_{d,(j-1)}-\hat{p}_{d,(j)}$ across the aperture, is denoted by $\hat{u}_{(j-1,j)}$. Similarly, $\hat{u}_{(j,j+1)}$ denotes the transverse acoustic velocity on $\sigma_{3,j}$.}}\label{Figure 2}
\end{figure}

 \zw{The $j^{\mathrm{th}}$ can is enclosed by the control volume $\mathcal{V}_{j}$ with boundary $\sigma_{j}$ (see Fig. \ref{Figure 2}). In the frequency domain, the Helmholtz equation and the corresponding BCs read \cite{Maling63}}
\begin{alignat}{2}
    \frac{\partial^2 \hat{p}_{(j)}(s,x)}{\partial x^2}-\left(\frac{s}{c_{j}}\right)^2\hat{p}_{(j)}(s,x)&=-s\frac{\gamma_{j}-1}{c_j^2}\hat{Q}_{(j)}(s,x)\quad&  \text{in $\mathcal{V}_{j}$},\label{Helmholtz eq.}\\
    \frac{\hat{p}_{(j)}(s,x)}{\hat{\boldsymbol{u}}_{(j)}(s,x)\cdot \boldsymbol{n}}&=Z_{(j)}(s,x)\quad& \text{on $\sigma_{j}$} \label{BC for Helmholtz eq.}.
\end{alignat}
In Eqs. \eqref{Helmholtz eq.} and \eqref{BC for Helmholtz eq.}, $\hat{p}_{(j)}$ and $\hat{\boldsymbol{u}}_{(j)}$ denote the acoustic pressure and velocity in the $j^{\mathrm{th}}$ can, $s=\nu+\mathrm{i} \omega$ is the Laplace variable, where $\omega$ and $\nu$ are the angular frequency and growth rate of thermoacoustic oscillations at a frequency $f=\omega/2\pi$, respectively, $\mathrm{i}$ is the imaginary unit, $\gamma_{j}$ and $c_j$ are the specific heat ratio and the ambient speed of sound in the $j^{\mathrm{th}}$ can, respectively, $\boldsymbol{n}$ is the outward facing normal vector to the boundary $\sigma_{j}$, $Z_{(j)}$ is the acoustic impedance on $\sigma_{j}$ and $\hat{Q}_{(j)}$ is the unsteady heat release rate fluctuations per unit \fc{volume in} the flame \fc{region}. By symmetry, we set $\gamma_{j}\equiv \gamma$ and $c_j \equiv c \hspace{0.2cm} \forall j$ in the following. 

\zw{The thermoacoustic model of the $j^{\mathrm{th}}$ can is sketched in Fig. \ref{Figure 2}. The dimensions are not true to scale. In Fig. \ref{Figure 2}\textbf{(a)}, $\mathcal{V}_{j}$ denotes the control volume, $A$ the cross-section area and $L$ the length of the can, respectively.} \ei{As discussed in $\S$\ref{Section: setting}, for simplicity, we assume a constant cross-section area along the can, because the details of the can geometry and acoustic-flame interactions are not in the scope of this study, which considers an idealized system.} \zw{$W$ is the width of the apertures between the cans, which are assumed to be rectangular with height $B$ (see Fig. \ref{Figure 1}). $\hat{Q}$ is the unsteady heat release rate across the flame front. The boundary $\sigma_j$ is divided into 4 parts: $\sigma_{1,j}$ at the can walls and burner outlet, $\sigma_{2,j}$ and $\sigma_{3,j}$ at the coupling interfaces, where the $j^{\mathrm{th}}$ can is connected to the $(j-1)^{\mathrm{th}}$ and $(j+1)^{\mathrm{th}}$ cans, respectively, and $\sigma_{4,j}$ at the turbine inlet at $\tilde{x}=L$. In Fig. \ref{Figure 2}, the acoustic velocity in streamwise direction and the acoustic pressure are denoted by $\hat{u}_{(j)}$ and $\hat{p}_{(j)}$, respectively, and $\hat{u}_{d,(j)}$ and $\hat{p}_{d,(j)}$ denote these quantities in the downstream section of each can, i.e., immediately upstream of the outlet of the cans:
\begin{align}
    \hat{u}_{d,(j)}(s)=&\,\hat{u}_{(j)}(s,\tilde{x}=L-W), \label{ds 1}\\
    \hat{p}_{d,(j)}(s)=&\,\hat{p}_{(j)}(s,\tilde{x}=L-W), \label{ds 2}
\end{align} 
with $L\gg W$, so that $W$ can be dropped from the argument on the RHS of Eqs. \eqref{ds 1} and \eqref{ds 2}, respectively. The transverse acoustic velocity on $\sigma_{2,j}$, which results from the pressure difference $\hat{p}_{d,(j-1)}-\hat{p}_{d,(j)}$ across the aperture, is denoted by $\hat{u}_{(j-1,j)}$. Similarly, $\hat{u}_{(j,j+1)}$ denotes the transverse acoustic velocity on $\sigma_{3,j}$. }

The parameter values used in the numerical examples throughout this work are listed in Table \ref{Table 1}. These values are in the range of those found in realistic H-class gas turbines.

\begin{table}[!ht]
\centering
\begin{tabular}{l l l}
\hline
\textbf{Parameter} & \textbf{Meaning} & \textbf{Value}\\
\hline
$N$ & Number of cans & $12$ \\
$W$ & Aperture width & $\{5,2\}$ cm\\
$B$ & Aperture height & $25$ cm \\
$A_a=WB$ & \zw{Cross-section area of the aperture }& $\{125,50\}$ cm$^2$ \\
$A$ & Cross-section area of the cans & $0.15$ m$^2$ \\
$L$ & Can length & $1.2$ m \\
$V=AL$ & Can Volume & $0.18$ m$^3$\\
\fc{$\omega_k$} & Natural eigenfrequency & $800$ rad/s\\
\fc{$\nu_0/\omega_k$} & \zw{Normalized base growth rate} & $\{3\%,-1.2\%\}$\\
\fc{$h/W$} & \zw{Normalized can spacing}& $\in[0,0.5]$\\
$U_\mathrm{tot}$ & \fc{Bulk velocity of combustion products} & $\{40,30\}$ m/s \\
$U=U_\mathrm{tot}/2$ & \fc{Vorticity disturbance advection speed} & $\{20,15\}$ m/s\\
$c$ & \zw{Ambient speed of sound} & $800$ m/s \\

\hline
\end{tabular}
\caption{Parameter values used in the numerical examples in this work. \label{Table 1}}
\end{table}

\zw{As discussed in $\S$\ref{Section: setting}, we restrict our analysis to low-frequency longitudinal eigenmodes whose wavelengths are large compared to the dimensions of the aperture $W$ and $B$. Since the boundaries $\sigma_{2,j}$, $\sigma_{3,j}$ are assumed to be compact with respect to the wavelength of the acoustic pressure oscillations in the can, the spatial dependence of the impedance $Z_{(j)}$ on these parts of the boundary can be neglected.}

\vi{We assume that the thermoacoustic dynamics in the cans are dominated by a single eigenmode $\psi_{k}$ with corresponding modal amplitude $\hat{\eta}_{(j),k}$ and eigenfrequency $\omega_k$. This assumption is expected to be satisfied in a frequency range around $f_k=\omega_k/2\pi$, and is confirmed by the acoustic pressure spectrograms from a real engine shown in Fig. 8 in \cite{ghirardo18}, where one can observe that the active modes are spread around $\pm 5\%$ of $f_k$.} Expanding the acoustic pressure in terms of $\psi_k$ yields
\begin{align}
    \hat{p}_{(j)}(s,x)=&\,\hat{\eta}_{(j),k}(s) \psi_{k}(x). \label{Modal expansion of complex pressure amplitude}\\
    \hat{\boldsymbol{u}}_{(j)}(s,x)=&- \frac{\hat{\eta}_{(j),k}(s) \boldsymbol{\nabla} \psi_{k}(x)}{s \rho}, \label{Modal expansion of complex velocity amplitude}
\end{align} 
where $\hat{\eta}_{(j),k}$ is the dominant modal amplitude defined by
\begin{align}
    \hat{\eta}_{(j),k}=&\,\frac{s \rho c^2}{s^2+\omega_k^2}\frac{1}{V_j\Lambda_j}\left(\frac{\gamma-1}{\rho c^2} \int_{\mathcal{V}_{j}} \hat{Q}_{(j)}(s,x)\psi_{k}(x)dV\right.\nonumber\\
    &-\left.\hat{\eta}_{(j),k}\int_{\sigma_{j}} \frac{|\psi_{k}(x)|^2}{Z_{(j)}(s,x)}dS\right),\quad j={1,...,N}, \label{Dominant modal amplitude}
\end{align} 
where $\Lambda_j=1/V_j \int_{\mathcal{V}_j} |\psi_k|^2 dV$ is the mode normalization factor and $V_j=\mathrm{Vol}(\mathcal{V}_j)$ is the volume of $\mathcal{V}_j$. By symmetry, $V_j\equiv V$, $\Lambda_j\equiv \Lambda$ and $Z_{(j)}\equiv Z\hspace{0.2cm}\forall j$. \ei{Equation \eqref{Dominant modal amplitude}, which is derived in the supplementary material, describes the projection of the Helmholtz equation \eqref{Helmholtz eq.} onto the eigenmode $\psi_k$ \cite{morse1986theoretical,NOIRAY20122753}.} \zw{It involves the acoustic impedance $Z_{(j)}$ at the boundary $\sigma_j$ as an unkown. }

\zw{The unimodal projection resulting in Eqs. \eqref{Modal expansion of complex pressure amplitude} and \eqref{Modal expansion of complex velocity amplitude} is performed under the assumptions that the system trajectories $p$ remain close to (a) the origin and (b) the linear eigenspace of the unforced Helmholtz equation spanned by $\psi_k$. These assumptions ensure that (a) the acoustic pressure signal is quasi-sinusoidal, which is a standard assumption of low-order thermoacoustic models \cite{Noiray16,Lieuwen03,culick2006unsteady,BONCIOLINI2021396}, and that (b) Eqs. \eqref{Modal expansion of complex pressure amplitude} and \eqref{Modal expansion of complex velocity amplitude} approximate well the acoustic pressure dynamics.}

\zw{In the following, we drop the subscript $k$ on $\eta_{(j),k}$. To arrive at an expression for $Z$, we use the Rayleigh conductivity, which is defined as follows \cite{HOWE1997,howe_1998,Howe1996}:
\begin{equation}
    K_R=-\frac{s \rho \hat{\Phi}}{[\hat{p}]} \label{Rayleigh conductivity for projected HH derivation}
\end{equation}
where $\hat{\Phi}$ is the outward facing coherent volume flux through the aperture, resulting from oscillatory motion of the vortex sheet, and $[\hat{p}]$ is the acoustic pressure difference across the aperture. By equating the coherent volume flux $\hat{\Phi}$ to the acoustic volume flux through the aperture $A_a  \hat{u}_a$, where $A_a=WB$ is the aperture area and $\hat{u}_a$ is the transverse acoustic velocity of the fluid in the aperture, we obtain a relation between $K_R$ and the specific acoustic impedance $Z_s=Z/\rho c$ \cite{jing1999experimental,SUN2002557,Tiemo20}:
\begin{equation}
    Z_s=\frac{[\hat{p}]}{\rho c \hat{u}_a}=-\frac{s A_a}{c K_R}. \label{Rayleigh conductivity}
\end{equation}
We define the heat release rate of the flame $\hat{q}_{(j)}$ as follows:}
\zw{\begin{equation}
 \hat{q}_{(j)}=\,\frac{\gamma-1}{V\Lambda}s\int_{\mathcal{V}_{j}}\hat{Q}_{(j)}(s,x)\psi_{k}(x)dV. \label{q}
 \end{equation}
 We model $\hat{q}_{(j)}$ as a linear function of the modal amplitude $\hat{\eta}_{(j)}$, which is justified for small enough acoustic pressure amplitudes (see, e.g., Fig. 2 in \cite{Noiray16}). Hence we write $\hat{q}_{(j)}=s\beta \hat{\eta}_{(j)}$, where $\beta$ is a real number describing the coherent flame response to acoustic perturbations. When $\beta$ is positive, the thermoacoustic feedback is constructive.}

\zw{We assume a mode normalization factor of $\Lambda=1/2$, which is exact for the longitudinal eigenmodes $\psi_k=\cos{(k \pi x/L )}$, $k\in\mathbb{N}$, of can combustors corresponding to the following limit case: $Z(s,x)\rightarrow \infty$ for $x\in\sigma_{1,j}$ and $W\rightarrow 0$. Because the coupling interfaces are acoustically compact, we set $dS =A_a \delta (\tilde{x}-L) d\tilde{x}$ on $\sigma_{2,j}$ and $\sigma_{3,j}$. As stated in $\S$\ref{Section: setting}, following \cite{YOON2020115774}, we assume a pressure antinode $\psi_k(\tilde{x}=L)=1$ at the turbine inlet.}

\ei{Under the above assumptions, following the steps detailed in the supplementary material, the projected Helmholtz equation \eqref{Dominant modal amplitude} can be rewritten as follows:}
\zw{\begin{equation}
    (s^2-2\nu_0 s+ \omega_k ^2) \hat{\eta}_{(j)}= s\varkappa(s)(\hat{\eta}_{(j-1)}+\hat{\eta}_{(j+1)}-2\hat{\eta}_{(j)}),\quad j={1,...,N} \label{Dynamics of dominant modal amplitude},
\end{equation}
where $\nu_0=(\beta-\alpha_0)/2$ is the thermoacoustic growth rate and we have defined the damping constant $\alpha_0$ and the frequency-dependent coupling term $\varkappa$ as follows:
\begin{align}
    \alpha_0=&\,\frac{2\rho c^2}{V}\int_{\sigma_{1,j}} \frac{|\psi_k(x)|^2}{Z(s,x)}\mathrm{d}S,\label{Alpha, no Bloch theory}\\
    \varkappa(s)=&-\frac{4 c^2 \mathcal{R} K_R(s)}{s V }.\label{Beta, no bloch theory}
\end{align}
In our low-order model, all dissipative effects at the boundary $\sigma_{1,j}$ are compounded into the damping constant $\alpha_0$, which, for simplicity, is assumed to be real and positive. For constructive thermoacoustic feedback, if $\beta>0$ exceeds $\alpha_0$, the growth rate $\nu_0$ becomes positive and an instability occurs \cite{NOIRAYSCHUERMANS2013152}. \cite{boujo2016quantifying} present a method to measure $\beta$ and $\alpha_0$ separately. In practice, $\nu_0$ depends on the operating condition parameters such as the equivalence ratio or the operating pressure. }
\zw{For $\omega\approx\omega_k$, the system of ODEs \eqref{Dynamics of dominant modal amplitude} describes the linear dynamics in the frequency domain of the dominant modal amplitudes $\hat{\eta}_{(j)}$, $j=1,...,N$, in an idealized can-annular combustor with $N$ cans.}

\subsection{Aeroacoustic coupling} \label{Section 3: Coupling}
\zw{In this section, following \cite{HOWE1997}, we derive the Rayleigh conductivity $K_R$ defined in Eq. \eqref{Rayleigh conductivity for projected HH derivation} which determines the frequency-dependent coupling term $\varkappa$ in Eq. \eqref{Dynamics of dominant modal amplitude}.}

\zw{For simplicity, we adopt the notation used in \cite{HOWE1997} with a complex angular frequency $\omega_c=\omega+\mathrm{i}\nu$. The forced hydrodynamic motion of the turbulent wake forming between neighboring cans is modeled as two vortex sheets separated by the can spacing $h$ which are subject to an oscillating pressure load $[p]e^{-\mathrm{i} \omega_c t}$, where $[p]=p_+-p_-$ and $p_\pm$ are the uniform pressure components on either side of the aperture. The vortex sheets separate two regions of constant mean axial velocity $U$. The pressure load causes a (complex) displacement of the vortex sheet $\zeta(\omega_c,\xi) e^{-\mathrm{i} \omega_c t}$ across the aperture, where $\zeta$ is the amplitude of the vortex sheet displacement and $\xi$ is a scaled streamwise variable defined as $\xi=2x/W-1$, which originates in the middle of the aperture and is equal to $\pm1$ at its edges. For compactness, the dependence of $\zeta$ on $\omega_c$ is suppressed below.} \ei{One finds that for $\omega\in\mathbb{R}$, $K_R$ depends only on the nondimensional Strouhal number 
\begin{equation}
    \mathrm{St}=\frac{\omega W}{2U},
\end{equation} 
which combines the acoustic oscillation frequency $\omega$ and the frequency of the hydrodynamic vorticity fluctuations in the turbulent wake $W/U$.}

\zw{By expressing the pressure perturbations on either side of the wake in terms of the velocity potentials $\phi_\pm$ and requiring the pressure on either side to be equal, Howe arrives at the following equation:
\begin{equation}
    p_+-\rho\left(-i\omega_c+U_+\frac{\partial}{\partial x}\right)\phi_+=p_--\rho\left(-i\omega_c+U_-\frac{\partial}{\partial x}\right)\phi_-+h\omega_c^2 \zeta, \label{Unsteady Bernoulli equation, not appendix}
\end{equation}
where $U_\pm$ are the axial mean flow speeds on either side of the aperture and $\phi_\pm$ are the velocity potentials associated with the velocity component normal to the aperture plane \cite{HOWE1997}. The last term on the RHS of \eqref{Unsteady Bernoulli equation, not appendix} accounts for the pressure difference induced by the finite thickness of the aperture $h$. Expressions for $\phi_\pm$ are given in Eq. (2.3) of \cite{HOWE1997}:
\begin{equation}
    \phi_\pm(x,z)=\mp\frac{1}{2\pi}\int_{0}^{W} \int_{0}^{B} \frac{ v_\pm(\bar{x}) }{\sqrt{(x-\bar{x})^2+(z-\bar{z})^2}}\mathrm{d}\bar{z}\mathrm{d}\bar{x}, \label{Velocity potential, original form}
\end{equation}
where $\bar{x}$ and $\bar{z}$ are integration variables corresponding to $x$ and $z$, respectively. The normal velocity just above and below the wake, $v_{\pm}$, is expressed in terms of $\zeta$:}
\zw{\begin{equation}
    v_{\pm}(x)=\left(-i\omega_c+U_\pm\frac{\partial}{\partial x}\right)\zeta(x).  \label{velocity in terms of zeta}
\end{equation}
Consistent with Refs. \cite{howe_1998,HOWE1997}, assuming strongly correlated fluid motion in spanwise direction $z$, we neglected the dependence of $\zeta$ on $z$ in Eqs. \eqref{Velocity potential, original form} and \eqref{velocity in terms of zeta}.}

\vi{Details of the derivation are explained in the supplementary material.} \zw{By combining Eq. \eqref{Velocity potential, original form} with Eq. \eqref{velocity in terms of zeta}, performing the integration over $\bar{z}$ in Eq. \eqref{Velocity potential, original form} and taking the average of Eq. \eqref{Unsteady Bernoulli equation, not appendix} over the spanwise direction $z$, Howe arrives at the following equation:
\begin{eqnarray}
    &&\int^1_{-1}\zeta'(\mu)\{\ln |\xi-\mu|+L_+(\xi,\mu)\}d\mu \nonumber\\
    &&\quad\quad\quad-\pi\mathrm{St}_c^2\left(\frac{h}{W}\right)\int^1_{-1}\zeta'(\mu)G(\xi,\mu)d\mu+(\lambda_+ +\lambda_- \xi) e^{\mathrm{i} \mathrm{St}_c \xi}=1, \label{Equation for zeta, full model}
\end{eqnarray}
 where $|\xi|<1$, $\zeta'=-\rho|\omega_c|^2 W \zeta/\pi [p]$, $\mathrm{St}_c=\omega_c W/2U$ is the Strouhal number based on the complex frequency $\omega_c$, $\mu$ is an integration variable corresponding to $\xi$, $\lambda_\pm$ are constants of integration,}
\zw{\begin{equation}
    G(\xi,\mu)=-H(\xi-\mu)(\xi-\mu)e^{\mathrm{i}\mathrm{St}_c(\xi-\mu)}, \label{Green's function, not appendix}
\end{equation}
where $H(\cdot)$ is the Heaviside function and}
\zw{\begin{eqnarray}
    L_+(\xi,\mu)=-\ln \{2B/W+\sqrt{(2B/W)^2+(\xi-\mu)^2}\}+\quad\quad\quad\nonumber\\
    \sqrt{1+(W/2B)^2(\xi-\mu)^2}-(W/2B)|\xi-\mu|.
\end{eqnarray}
\dr{Equation \eqref{Equation for zeta, full model} can be understood as a condition for the spanwise average pressure continuity across the vortex sheet \cite{Howe1996} and coincides, up to a typographical error (a factor 2 before the second integral), with Eq. (2.11) in \cite{HOWE1997}. Note that for comparison, the terms involving $\lambda_\pm$ need to be replaced according to the remark on p. 356 in the latter reference.}}

\zw{We seek the solution $\zeta'(\xi)$, $\xi\in[-1,1]$, of Eq. \eqref{Equation for zeta, full model} satisfying the Kutta condition, which states that the vortex sheet leaves the upstream edge leave the upstream edge smoothly \cite{howe_1981}:}
\zw{\begin{equation}
    \zeta'(-W/2)=\frac{\partial \zeta' }{\partial x}(-W/2)=0. \label{Kutta condition}
\end{equation}
From this solution $\zeta'$, using $\Phi=i\omega_c \int^W_0 \int^B_0 \zeta dx dz$ and the fact that $\hat{\Phi}/[\hat{p}]=\Phi/[p]$, the Rayleigh conductivity \eqref{Rayleigh conductivity for projected HH derivation} can be computed from the following formula \cite{HOWE1997}:}
\zw{\begin{equation}
    K_R(\omega_c)=-\frac{\pi B}{2} \int^1_{-1}\zeta'(\mu,\omega_c)d\mu. \label{KR from Howe model}
\end{equation}
To obtain $K_R(s)$, one has to evaluate }
\zw{\begin{equation}
    K_R(s)=K_R(\omega_c^*),
\end{equation}
where $(\cdot)^*$ denotes the complex conjugate. The conjugate of $\omega_c$ appears because of different conventions in the definitions of $\omega_c$ and the Laplace variable $s=\mathrm{i}\omega_c^*$. Equation \eqref{Equation for zeta, full model} is an integral equation which is here solved numerically using Gauss-Legendre quadrature with 15 ($\S$\ref{Subsection parameter study}) or 40 nodes ($\S$\ref{Subsection discussion}). Details of the numerical method used to solve Eq. \eqref{Equation for zeta, full model} are discussed in the supplementary material.}

\zw{For large aspect ratios $B/W\gg 1$ and vanishing wall thickness $h/W\ll 1$, the following formula for the thin-wall approximation of $K_R$ can be derived \cite{HOWE1997,howe_1981}:
\begin{equation}
    K_R(\omega_c)=\frac{\pi B}{2\big[F(\mathrm{St}_c)+\ln (8B/\mathrm{e}W)\big]}, \label{KR, thin wall, infinite b}
\end{equation}
where $\mathrm{e}$ is Euler's number and
\begin{equation}
    F(\mathrm{St}_c)=\frac{J_0(\mathrm{St}_c)K(\mathrm{St}_c)-\big[J_0(\mathrm{St}_c)-2K(\mathrm{St}_c)\big]M(\mathrm{St}_c)}{\mathrm{St}_c\big(J_0(\mathrm{St}_c)J_1(\mathrm{St}_c)+\mathrm{St}_c\{J_1(\mathrm{St}_c)^2+[J_0(\mathrm{St}_c)-2 \mathrm{i} J_1(\mathrm{St}_c)]^2\}\big)}, \label{F from Howe}
\end{equation}
where $K(x)=\mathrm{i} x \left[J_0(x)-\mathrm{i}J_1(x) \right]$, $M(x)=\left[J_0(x)-\mathrm{i}x(J_0(x)+ \mathrm{i} J_1(x)\right]$ and $J_0$ and $J_1$ are Bessel functions of the first kind \cite{bowman2012introduction}. 
In the case of vanishing mean flow, $U\equiv 0$. For the thin-wall approximation \eqref{KR, thin wall, infinite b}, this implies $F\equiv0$ \cite{Howe1996} and
\begin{equation}
    K_R=\frac{\pi B}{2\ln (8B/\mathrm{e}W)}. \label{KR, no-flow}
\end{equation}
}

\begin{figure}[!t]
\begin{psfrags}
\psfrag{a}{$0$}
\psfrag{b}{$2$}
\psfrag{c}{$4$}
\psfrag{d}{$6$}
\psfrag{e}{$8$}
\psfrag{f}{$10$}
\psfrag{g}{$1$}
\psfrag{h}{$0.5$}
\psfrag{i}{$0$}
\psfrag{j}{$-0.5$}
\psfrag{k}{$0$}
\psfrag{l}{$-0.2$}
\psfrag{m}{$-0.4$}
\psfrag{n}{$-0.2$}
\psfrag{o}{$-0.8$}
\psfrag{p}{$-0.2$}
\psfrag{q}{$-0.4$}
\psfrag{r}{$-0.6$}
\psfrag{s}{$1.4$}
\psfrag{G}{$1.5$}
\psfrag{S}{$-1$}
\psfrag{A}{$2.8$}
\psfrag{B}{$2$}
\psfrag{C}{$1.2$}
\psfrag{D}{$0.4$}
\psfrag{E}{$-0.4$}
\psfrag{F}{$1.6$}
\psfrag{G}{$1$}
\psfrag{H}{$0.4$}
\psfrag{I}{$-0.4$}
\psfrag{t}{\hspace{0.07cm}$\omega W / 2U$}
\psfrag{u}{\hspace{0.28cm}$\mathrm{Re}(K_R/B)$}
\psfrag{v}{\hspace{0.26cm}$\mathrm{Im}(K_R/B)$}
\psfrag{W}{\textcolor{krgreen}{$0.1$}}
\psfrag{w}{b}
\psfrag{X}{\textcolor{krorange}{$0.5$}}
\psfrag{W}{\textcolor{krblue}{$h/W=0$}}
\psfrag{Y}{\textcolor{krred}{\begin{tabular}{@{}c@{}}
no- \\
flow 
\end{tabular}}}
\psfrag{Z}{$h/W$}
\psfrag{M}{\hspace{-0.04cm}\textcolor{krgreen}{$0.02$}}
\psfrag{T}{\textcolor{krpurple}{$0.06$}}
\psfrag{1}{$0$}
\psfrag{2}{$0.1$}
\psfrag{3}{\textcolor{krcyan}{$0.2$}}
\centerline{\includegraphics{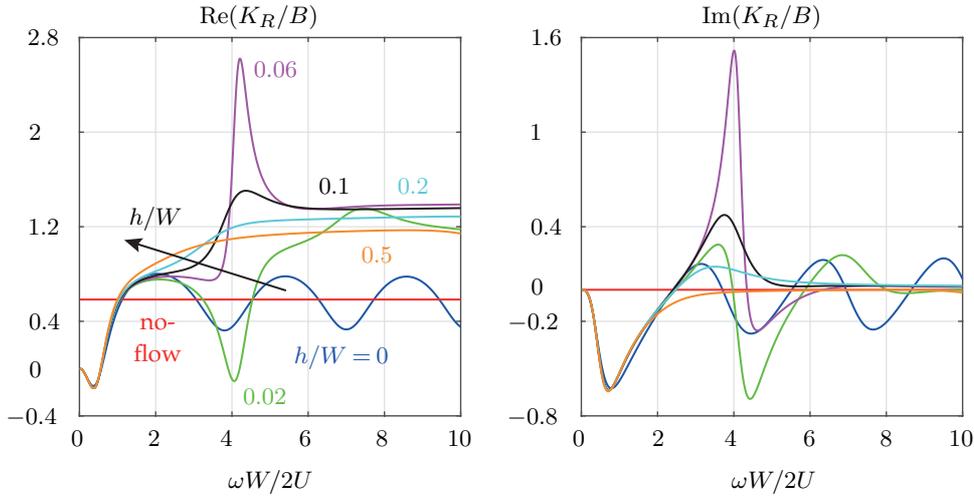}}
\end{psfrags}
\caption{\dr{Real and imaginary parts of the normalized Rayleigh conductivity $K_R/B$ as a function of the Strouhal number $\mathrm{St}=\omega W/2U\in[0,10]$, $\omega\in\mathbb{R}$ for different values of the can spacing $h/W\in\{0,0.02,0.06,0.1,0.2,0.5\}$ and aspect ratio $B/W=5$. The arrow indicates the direction of increasing can spacing $h/W$. The blue curve corresponds to the thin-wall approximation \eqref{KR, thin wall, infinite b}. For $h/W>0$, the curves are computed from Eq. \eqref{Equation for zeta, full model}. The no-flow limit of the thin-wall approximation \eqref{KR, no-flow} is shown in red. Regions where $\mathrm{Im}(K_R)>0$ indicate amplification the sound field by the acoustic-hydrodynamic interaction. } }\label{Figure 3}
\end{figure}
\zw{The real and imaginary parts of the normalized Rayleigh conductivity $K_R/B$ are plotted in Fig. \ref{Figure 3} as a function of the Strouhal number $\mathrm{St}=\omega W/2U\in[0,10]$, $\omega\in\mathbb{R}$ for different values of the can spacing $h/W\in\{0,0.02,0.06,0.1,0.2,0.5\}$ and aspect ratio $B/W=5$. The blue curve corresponds to the thin-wall approximation \eqref{KR, thin wall, infinite b}. For $h/W>0$, the curves are computed from Eq. \eqref{Equation for zeta, full model}. The no-flow limit of the thin-wall approximation \eqref{KR, no-flow} is shown in red. Regions where $\mathrm{Im}(K_R)>0$ indicate amplification the sound field by the mean flow. Indeed, it is straightforward to show using Eq. \eqref{Rayleigh conductivity} that $\mathrm{Im}(K_R)>0$ is equivalent to a reflection coefficient $R=(Z_s-1)/(Z_s+1)$ of the aperture with magnitude $|R|>1$, meaning that incident acoustic waves are reflected with an increased amplitude \cite{Tiemo20}. We note that for the parameter values listed in Table \ref{Table 1}, we have $B/W\geq5$, so that the assumption of a large aspect ratio in the derivation of the thin-wall approximation is roughly satisfied. }

\zw{We see in Fig. \ref{Figure 3} that, when the can spacing $h/W$ is increased, the amplification, measured by the maximum of $\mathrm{Im}(K_R)$, first increases and then decreases until around $h/W\approx 0.5$, the amplification is fully suppressed. For higher values of $h/W$, the acoustic-hydrodynamic interaction is purely dissipative. This is consistent with the study presented in Fig. 3 in \cite{HOWE1997}, albeit for a different value of the aspect ratio $B/W$.}

\dr{We see in Fig. \ref{Figure 3} that the thin-wall approximation in Eq. \eqref{KR, thin wall, infinite b} does not converge to its no-flow limit for $\mathrm{St}\rightarrow\infty$. This leads to the question of up to which value of the Strouhal number the thin-wall approximation can be considered a qualitatively correct representation of the reflection coefficient $R$ of the aperture under turbulent grazing flow. In previous work by our group on a similar configuration, impedance measurements have been presented over a frequency range where, on a part of this range, $|R|$ exceeds $1$ \cite{bourquard_faure-beaulieu_noiray_2021}. If such measurements are available, one can fit numerically the first undulation (damping at low Strouhal number and the first amplification region) to the experiments to obtain a physics-based quantitative model of the aperture's reflection coefficient \cite{Tiemo20}. Beyond the first undulation, the physical meaning of the thin-wall approximation \eqref{KR, thin wall, infinite b} is unclear, because it predicts the same repeating behavior, alternating between dissipative to amplifying for all Strouhal numbers. This implies a rough limit of validity of the thin-wall approximation \eqref{KR, thin wall, infinite b} at around $\mathrm{St}=4$, because this is where the first undulation in the imaginary part of $K_R$ ends. This means that the model is only valid for sufficiently high velocities (or sufficiently small apertures/frequencies), and will break down, as we decrease $U$, before we reach the no-flow limit, which therefore deserves a separate discussion. There is no such a priori limit of validity for the general Rayleigh conductivity model with $h/W>0$. The curves in Fig. \ref{Figure 3} suggest that the general model predicts the same qualitative behavior in the no-flow limit as Eq. \eqref{KR, no-flow}, namely that the acoustic-hydrodynamic interaction is purely reactive in that limit. }

\dr{Despite the shortcomings of the thin-wall approximation discussed above, using the analytical formula in Eq. \eqref{KR, thin wall, infinite b} significantly reduces the computational cost of (repeatedly) computing the frequency-dependent coupling term $\varkappa$ in Eq. \eqref{Beta, no bloch theory} compared to using the numerical solution $K_R$ of the integral equation \eqref{Equation for zeta, full model}. As we see in Fig. \ref{Figure 3}, for $\mathrm{St}\leq4$, the thin-wall approximation and the general model are qualitatively similar up to a wall thickness of $h/W\approx 0.02$. In the present study, we consider the aeroacoustic interaction of low-frequency thermoacoustic modes through compact apertures under turbulent grazing flow, which typically occurs at low to moderate Strouhal numbers $\mathrm{St}<4$, so that using the thin-wall approximation \eqref{KR, thin wall, infinite b} is justified for small enough $h/W$. To study the influence of the can spacing $h$ on the frequency spectrum, the numerical solution of Eq. \eqref{Equation for zeta, full model} is required.}

\subsection{Bloch wave ansatz} \label{Section 4: Bloch wave theory}
\zw{In this section, we use a Bloch wave ansatz to simplify the system of ODEs \eqref{Dynamics of dominant modal amplitude}, which describes the linear dynamics in the frequency domain of the dominant modal amplitudes $\hat{\eta}_{(j)}$, $j=1,...,N$. Following \cite{mensah2016efficient}, we assume that the acoustic pressure in the cans is a Bloch wave. In the present context, this means it is an eigenfunction of the translation operator $\mathrm{T}[\cdot]$, which is defined by}
\zw{\begin{equation}
    \mathrm{T}\, [\hat{p}_{(j)}]=\hat{p}_{(j+1)}.
\end{equation}
By making use of the general statement derived in \cite{mensah2016efficient}, \cite{ghirardo18} show that if $\hat{p}_{(j)}$ is a Bloch-wave, it can be expressed as }
\zw{\begin{equation}
    \hat{p}_(j)(s,x)=\Psi(s,x) e^{\mathrm{i}\theta b},
\end{equation}
where $b$ is the Bloch wavenumber, $\theta=-2\pi j/N$ is the discrete azimuthal coordinate along the ring of can combustors and $\Psi(s,x)$ is the same in every can. The minus sign appears because we use a different convention for the can order than \cite{ghirardo18}. }

\zw{In the present work, the quantity of interest is the downstream acoustic pressure $\hat{p}_{d,(j)}$, which is spatially constant due to the assumption of acoustically compact coupling apertures. Indeed, with the unimodal expansion \eqref{Modal expansion of complex pressure amplitude}, it can be written as $\hat{p}_{d,(j)}(s)=\hat{\eta}_{(j)}(s)\psi_k(\tilde{x}=L)$. Using $\psi_k(\tilde{x}=L)=1$, the Bloch wave ansatz simplifies to 
\begin{equation}
    \hat{\eta}_{(j)}(s)=\Psi(s) e^{\mathrm{i}\theta b}\quad\forall j, \label{Simplified Bloch wave ansatz}
\end{equation}
where $\Psi\in\mathbb{C}$ is spatially constant. Different values of $b$ correspond to different azimuthal phase patterns along the turbine annulus \cite{ghirardo18}, which we call Bloch modes in the following.}

\begin{figure}[!t]
\begin{psfrags}
\psfrag{a}{\hspace{0.1cm}$\arg\hat{\eta}_{(j)}$}
\psfrag{b}{(b)}
\psfrag{c}{c}
\psfrag{d}{d}
\psfrag{e}{e}
\psfrag{f}{f}
\psfrag{g}{g}
\psfrag{h}{h}
\psfrag{i}{0}
\psfrag{j}{$\pi$}
\psfrag{k}{$2\pi$}
\psfrag{l}{\hspace{0.1cm}$b=0$}
\psfrag{m}{\hspace{0.1cm}$b=1$}
\psfrag{n}{\hspace{0.1cm}$b=2$}
\psfrag{o}{\hspace{0.1cm}$b=3$}
\psfrag{p}{\hspace{0.1cm}$b=4$}
\psfrag{q}{\hspace{0.1cm}$b=5$}
\psfrag{r}{\hspace{0.1cm}$b=6$}
\centerline{\includegraphics{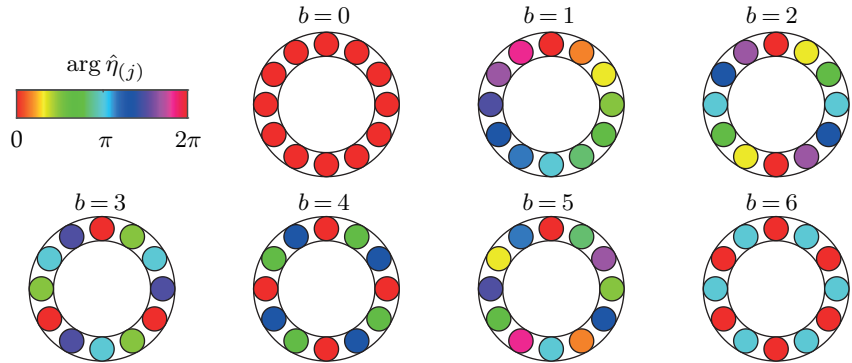}}
\end{psfrags}
\caption{\zw{All possible distinct Bloch modes with non-negative Bloch wavenumber $b$ in a ring of $N=12$ oscillators. The color bar indicates the value of the phase of the modal amplitude $\hat{\eta}_{(j)}$. The respective Bloch modes for negative $b$ can be obtained by reversing the can order.}}\label{Figure 4}
\end{figure}

\zw{We visualize all possible distinct Bloch modes with non-negative $b$ for $N=12$ in Fig. \ref{Figure 4}. The color bar indicates the value of the phase of the modal amplitude $\hat{\eta}_{(j)}$. The respective Bloch modes for negative $b$ can be obtained by reversing the can order. }

\zw{Using Eq. \eqref{Simplified Bloch wave ansatz}, we express the modal amplitudes of neighboring cans as follows:
\begin{equation}
    \hat{\eta}_{(j+1)}=\hat{\eta}_{(j)} e^{-\mathrm{i}\frac{ 2\pi b}{N}} \quad \text{and} \quad \hat{\eta}_{(j-1)}=\hat{\eta}_{(j)} e^{\mathrm{i}\frac{ 2\pi b}{N}}, \label{Bloch boundary conditions}  
\end{equation}
where $b\in \left]\mathrm{ceil}\left(-N/2\right),\mathrm{floor}\left(N/2\right)\right]$. Substituting Eq. \eqref{Bloch boundary conditions} into Eq. \eqref{Dynamics of dominant modal amplitude} and assuming a nonzero perturbation $\eta_{(j)}\neq 0$ yields}
\zw{\begin{equation}
    s^2-\big[2\nu_0 -4\varkappa(s)\sin^2(\pi b/N) \big]s+ \omega_k ^2=0,\quad j={1,...,N} \label{Dynamics of dominant modal amplitude, bloch theory},
\end{equation}
where the trigonometric identity $1-\cos{x}=2\sin^2{(x/2)}$ was used. The complex solutions $s$ of equation \eqref{Dynamics of dominant modal amplitude, bloch theory} describe the frequency spectrum $(\omega,\nu)$ of our model. Because $\sin^2(\cdot)$ is an even function, the spectrum is degenerate with respect to positive and negative values of $b$.}

\zw{In Eq. \eqref{Dynamics of dominant modal amplitude, bloch theory}, the coupling between neighboring cans is now described implicitly in terms of the the Bloch wavenumber $b$. For computational purposes, it is useful to rewrite Eq. \eqref{Dynamics of dominant modal amplitude, bloch theory} as follows:
\begin{equation}
    s^2-2 \nu_0 s +\omega_{k} ^2- b_0 K_R(s)\sin ^2(\pi b/N)=0 \label{Frequency spectrum, effect of mean flow},
\end{equation}
where $b_0=16 \mathcal{R} c^2/ V>0 $. In this work, Eq. \eqref{Frequency spectrum, effect of mean flow} was solved numerically for $s=\mathrm{i}\omega+\nu$ using the $\textit{fsolve}$ function with default options in MATLAB 2020b \cite{MATLAB:2020}.} 

\section{Results \label{Section 5: Results}}
\subsection{Parameter study\label{Subsection parameter study}}
\zw{In this section, we perform a parameter study on the frequency spectrum of the thermoacoustic model derived in $\S$\ref{Section 2: Thermoacoustics}. To do this, we vary different parameters and repeatedly solve Eq. \eqref{Frequency spectrum, effect of mean flow} at each point.}

\zw{As stated in $\S$\ref{Section: setting}, we restrict ourselves to the study of small perturbations of the frequency spectrum by the coupling. By the implicit function theorem, if $x_0$ is a zero of a function $f(x)$, then for $\partial f/\partial x|_{x=x_0} \neq0$ and small enough $\varepsilon>0$, the perturbed function $f(x)+\varepsilon f_1(x)$ has a zero $x_0+\varepsilon x_1$ near $x_0$. Hence, for a small enough perturbation $b_0 K_R(s)\sin ^2(\pi b/N)$, there exists, given $\omega_k\neq\nu_0$, a solution of Eq. \eqref{Frequency spectrum, effect of mean flow} near the unperturbed solution $s_0=\nu_0+\mathrm{i}\sqrt{\omega_{k} ^2-\nu_0^2}$, which corresponds to a thermoacoustic instability of an isolated can. In this work, we focus on these perturbed solutions and do not consider other solutions that may emerge far away from $s_0$ from the zeros of $K_R$. }

\begin{figure}[!t]
\begin{psfrags}
\psfrag{a}{$0.05$}
\psfrag{b}{$0$}
\psfrag{c}{$-0.05$}
\psfrag{d}{$-0.1$}
\psfrag{e}{$0$}
\psfrag{f}{$1$}
\psfrag{g}{$2$}
\psfrag{h}{$3$}
\psfrag{i}{$4$}
\psfrag{j}{$1.2$}
\psfrag{k}{$1.1$}
\psfrag{l}{$1$}
\psfrag{m}{$0.9$}
\psfrag{n}{$0.1$}
\psfrag{o}{$0$}
\psfrag{p}{$-0.1$}
\psfrag{q}{$-0.2$}
\psfrag{r}{$2$}
\psfrag{s}{$2.5$}
\psfrag{t}{$3$}
\psfrag{u}{$3.5$}
\psfrag{v}{$4$}
\psfrag{w}{$1$}
\psfrag{x}{$0.9$}
\psfrag{y}{$0.8$}
\psfrag{z}{$0.7$}
\psfrag{A}{$0$}
\psfrag{B}{$-0.04$}
\psfrag{C}{$-0.08$}
\psfrag{D}{$1$}
\psfrag{E}{$0.9$}
\psfrag{F}{$0.8$}
\psfrag{G}{$\nu/\omega_k$}
\psfrag{H}{$\omega/\omega_k$}
\psfrag{I}{\hspace{0.35cm}Norm. growth rate }
\psfrag{J}{\hspace{0.35cm}Norm. \dr{frequency} }
\psfrag{K}{Unstable}
\psfrag{L}{Stable}
\psfrag{M}{\hspace{0.3cm}\begin{tabular}{@{}c@{}}
Instability \\
suppressed
\end{tabular}}
\psfrag{N}{\hspace{0.3cm}\begin{tabular}{@{}c@{}}
Coupling-induced \\
instability
\end{tabular}}
\psfrag{O}{\hspace{0.04cm} \begin{tabular}{@{}c@{}}
Dissipative \\
coupling
\end{tabular}}
\psfrag{P}{\textbf{(a)}}
\psfrag{Q}{\textbf{(b)}}
\psfrag{R}{\textbf{(c)}}
\psfrag{S}{\textbf{(d)}}
\psfrag{T}{\textbf{(e)}}
\psfrag{U}{\textbf{(f)}}
\psfrag{V}{$\mathrm{St}_k$}
\psfrag{W}{$b$}

\includegraphics[width=0.9\textwidth,center]{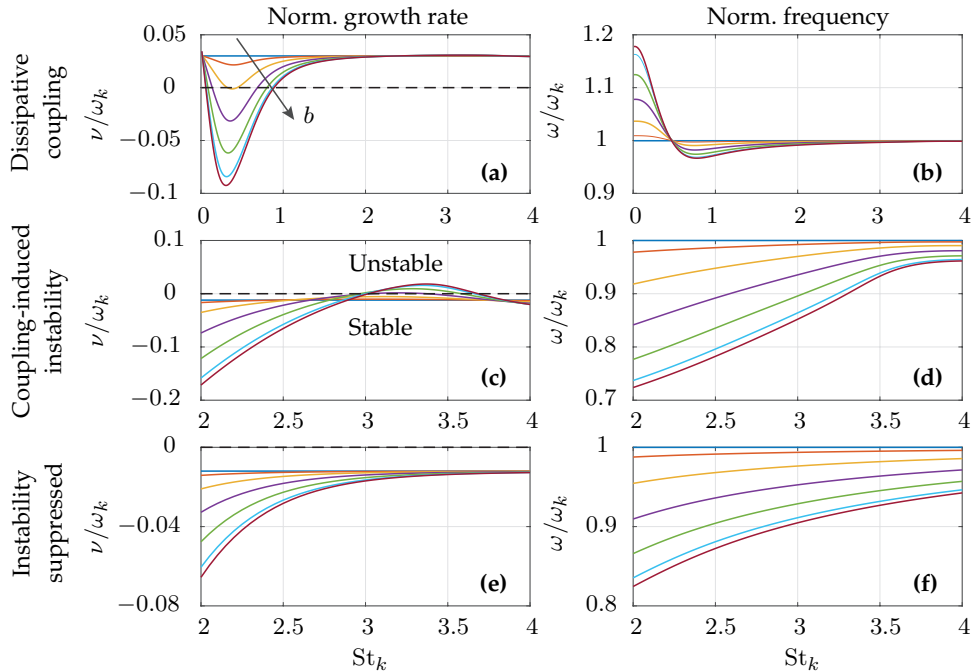}
\end{psfrags}
\caption{\zw{Frequency spectrum $(\omega,\nu)$ as a function of the normalized eigenfrequency $\mathrm{St}_k=\omega_k W / 2U$. The dashed black line marks the stability border $\nu=0$. \ei{The arrow indicates the direction of increasing Bloch wavenumber $b$.} In \textbf{(a)} and \textbf{(b)}, for the first set of parameters in Table \ref{Table 1}, we assume for each $\omega_k$ an unstable mode with positive base growth rate $\nu_0$ equal to $3\%$ of $\omega_k$. The insets \textbf{(c)}, \textbf{(d)}, \textbf{(e)} and \textbf{(f)} correspond to the second set of parameters in Table \ref{Table 1}, where we assumed stable modes with negative base growth rate $\nu_0$ equal to $-1.2\%$ of $\omega_k$. Vanishing can spacing $h/W$ was assumed in \textbf{(a)}-\textbf{(d)}, while in \textbf{(e)} and \textbf{(f)}, a finite value $h/W=0.5$ was used.}}\label{Figure 5}
\end{figure}

\zw{We begin by studying the influence of the natural eigenfrequency $\omega_k$ on the frequency spectrum in Fig. \ref{Figure 5}.} \ei{Note that there can be several longitudinal eigenmodes for which the aperture remains compact, and their eigenfrequency will depend on the speed of sound, the can length, the impedance BCs and the mode order. Therefore it is not straightforward to give a general expression of this eigenfrequency as a function of the physical parameters and we decide to simply \textit{assume} there exists a longitudinal mode with a given $\omega_k$ and base growth rate $\nu_0$. We compute the frequency spectrum $(\omega,\nu)$ for this mode as a function of the Bloch wavenumber $b$. If a (stable or unstable) mode with mode shape $\psi_k$ and spectrum $(\omega_k,\nu_0)$ does exist, this tells us whether or not, under perturbation by the aeroacoustic coupling, this mode is linearly stable and at what frequency the system oscillates.}

\zw{In Fig. \ref{Figure 5}\textbf{(a)} and \textbf{(b)}, we show the frequency spectrum $(\omega,\nu)$ as a function of the normalized eigenfrequency $\mathrm{St}_k=\omega_k W / 2U$ for the first set of parameters in Table \ref{Table 1} and vanishing can spacing $h/W$. \ei{In this and in the following figures, the dashed black line marks the stability border $\nu=0$ and the arrow indicates the direction of increasing Bloch wavenumber $b$.} For each $\omega_k$, we assume an unstable mode with positive base growth rate $\nu_0$ equal to $3\%$ of $\omega_k$. Figure \ref{Figure 5}\textbf{(a)} shows that at low  values of $\mathrm{St}_k$, the coupling has a dissipative effect, effectively damping some Bloch modes over a range of $\omega_k$ around $\mathrm{St}_k\approx 0.4$. In Fig. \ref{Figure 5}, we see that the eigenfrequency $\omega$ of the Bloch modes is increased by the coupling until about $\mathrm{St}_k\approx0.5$ and then decreased for higher values of $\mathrm{St}_k$. The effect of the coupling diminishes with increasing $\mathrm{St}_k$, and only negligible effects are observed above $\mathrm{St}_k=2$.} 

\ei{In Fig. \ref{Figure 5}\textbf{(c)} and \textbf{(d)}, we show the frequency spectrum $(\omega,\nu)$ for the second set of parameters in Table \ref{Table 1} with vanishing can spacing $h/W$. We assume stable modes with negative base growth rate $\nu_0$ equal to $-1.2\%$ of $\omega_k$. We see in Fig. \ref{Figure 5}\textbf{(c)} that the coupling makes some Bloch modes unstable around  $\mathrm{St}_k\approx 3.4$. Figure \ref{Figure 5}\textbf{(d)} shows that in the domain shown, the frequency $\omega$ is strictly decreased by the coupling.}

\zw{Going from Fig. \ref{Figure 5}\textbf{(c)} and \textbf{(d)} to Fig. \ref{Figure 5}\textbf{(e)} and \textbf{(f)}, the can spacing is changed from $h/W=0$ to a finite value of $h/W=0.5$, while all other parameters are unchanged. We observe that the coupling-induced instability shown in \ref{Figure 5}\textbf{(e)} is completely suppressed by the increased can spacing, while the frequency curves $\omega(\omega_k)$ shown in Fig. \ref{Figure 5}\textbf{(f)} remain qualitatively similar to those in Fig. \ref{Figure 5}\textbf{(d)}.} 

\zw{We see in Fig. \ref{Figure 5} that the push-push mode with $b=0$ is unaffected by the coupling and that higher-order Bloch modes are more strongly affected by the coupling than lower-order ones, which is expected from Eq. \eqref{Frequency spectrum, effect of mean flow}.}

\begin{figure}[!t]
\begin{psfrags}
\psfrag{a}{$\nu/\omega_k$}
\psfrag{b}{$\omega/\omega_k$}
\psfrag{c}{$\nu$}
\psfrag{d}{$\omega$}
\psfrag{e}{$\mathrm{St}_k$}
\psfrag{f}{$U_\mathrm{tot}$ [m/s]}
\psfrag{g}{$W$ [cm]}
\psfrag{h}{$\nu_0/\omega_k$}
\psfrag{k}{$1.2$}
\psfrag{l}{$1.1$}
\psfrag{m}{$1$}
\psfrag{n}{$0.9$}
\psfrag{o}{$-0.2$}
\psfrag{p}{$-0.1$}
\psfrag{q}{$0$}
\psfrag{r}{$0.1$}
\psfrag{u}{$4$}
\psfrag{t}{$3$}
\psfrag{s}{$2$}
\psfrag{m}{$1$}
\psfrag{P}{$b$}
\psfrag{v}{$0.8$}
\psfrag{y}{$40$}
\psfrag{x}{$80$}
\psfrag{w}{$120$}
\psfrag{z}{$0.6$}
\psfrag{A}{$0.4$}
\psfrag{C}{$-0.4$}
\psfrag{B}{$-0.6$}
\psfrag{D}{$6$}
\psfrag{E}{$4$}
\psfrag{F}{$2$}
\psfrag{G}{$850$}
\psfrag{H}{$800$}
\psfrag{I}{$750$}
\psfrag{J}{$700$}
\psfrag{K}{$-400$}
\psfrag{L}{$-200$}
\psfrag{N}{$-0.3$}
\psfrag{O}{$0.7$}
\psfrag{F}{$2$}
\psfrag{Z}{$0.3$}
\psfrag{Q}{\textbf{(a)}}
\psfrag{R}{\textbf{(b)}}
\psfrag{S}{\textbf{(c)}}
\psfrag{T}{\textbf{(d)}}
\psfrag{U}{\textbf{(e)}}
\psfrag{V}{\textbf{(f)}}
\includegraphics[width=0.85\textwidth,center]{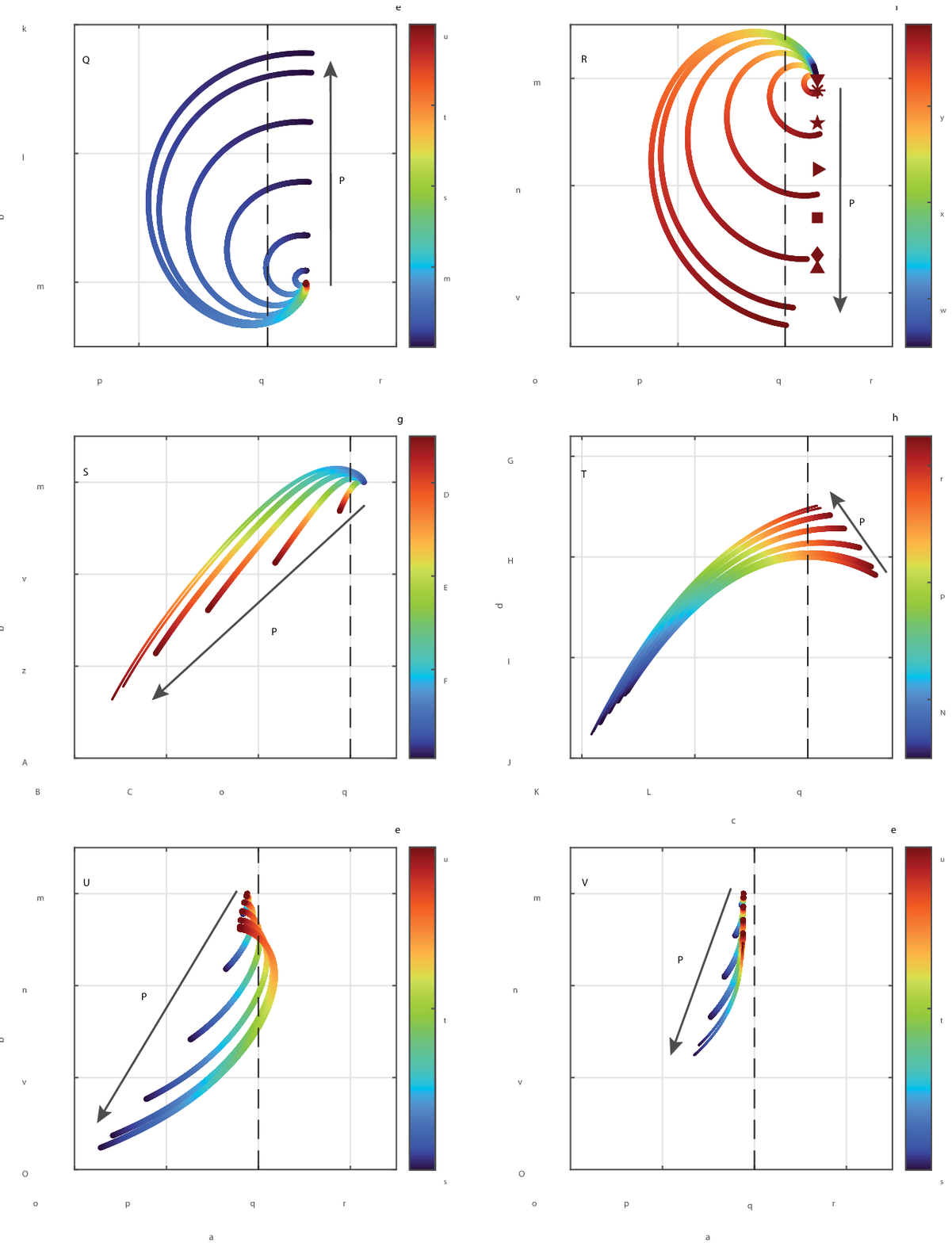}
\end{psfrags}
\caption{\zw{Parameter study in the root locus plane. The dashed black line marks the stability border $\nu=0$. \ei{The arrows indicate the direction of increasing Bloch wavenumber $b$.} For the first set of parameters in Table \ref{Table 1}, the frequency spectrum $(\omega,\nu)$ is plotted as a function of \textbf{(a)} the normalized eigenfrequency $\mathrm{St}_k=\omega_k W / 2U$, \textbf{(b)} the bulk velocity of the combustion products $U_\mathrm{tot}$, \textbf{(c)} the aperture with $W$ and \textbf{(d)} the normalized base growth rate $\nu_0/\omega_k$. \ei{The red symbols in \textbf{(b)} mark the no-flow limit of the thin-wall approximation \eqref{KR, no-flow}.} In \textbf{(e)} and \textbf{(f)}, the spectrum $(\omega,\nu)$ is plotted as a function of $\mathrm{St}_k$ for the second set of parameters in Table \ref{Table 1}. Going from Fig. \ref{Figure 6}\textbf{(e)} to \textbf{(f)}, the can spacing is increased from $h/W=0$ to $h/W=0.5$.}}\label{Figure 6}
\end{figure}

\zw{A parameter study in the root locus plane is presented in Fig. \ref{Figure 6}, where the frequency spectrum $(\omega,\nu)$ is plotted for the first set of parameters in Table \ref{Table 1} as a function of \textbf{(a)} the normalized eigenfrequency $\mathrm{St}_k=\omega_k W / 2U$, \textbf{(b)} the bulk velocity of the combustion products $U_\mathrm{tot}$, \textbf{(c)} the aperture with $W$ and \textbf{(d)} the normalized base growth rate $\nu_0/\omega_k$. In Fig. \ref{Figure 6}\textbf{(e)} and \textbf{(f)}, the spectrum $(\omega,\nu)$ is plotted as a function of the $\mathrm{St}_k$ for the second set of parameters in Table \ref{Table 1}. Going from Fig. \ref{Figure 6}\textbf{(e)} to \textbf{(f)}, the can spacing is increased from $h/W=0$ to $h/W=0.5$. The insets in Fig. \ref{Figure 6}\textbf{(a)}, \textbf{(e)} and \textbf{(f)} correspond to Fig. \ref{Figure 5}\textbf{(a)} and \textbf{(b)}, Fig. \ref{Figure 5}\textbf{(c)} and \textbf{(d)} and Fig. \ref{Figure 5}\textbf{(e)} and \textbf{(f)}, respectively. }

\ei{The red symbols in Fig. \ref{Figure 6}\textbf{(b)} mark the no-flow limit of the thin-wall approximation \eqref{KR, no-flow},} \zw{for which Eq. \eqref{Frequency spectrum, effect of mean flow} has the following exact solution:
\begin{equation}
    s_{1,2}(b)=\nu_0 \pm \mathrm{i} \sqrt{\omega_{k} ^2-\omega_b^2 \sin ^2(\pi b/N)-\nu_0^2}, \label{Solution for frequency spectrum}
\end{equation}
where $\omega_b^2=8 \pi c^2 \mathcal{R}  B/ V \ln (8B/\mathrm{e}W)>0$ and only the solution branch with positive imaginary part is considered.} \ei{Equation \eqref{Solution for frequency spectrum} implies that in the case of zero mean flow, the coupling between the cans is purely reactive, altering the reduced frequency $\sqrt{\omega_{k} ^2-\nu_0^2}$ of a single can but leaving the growth rate $\nu_0$ unchanged. If we set $\beta\equiv0$, this is consistent with the results shown in Fig. 11 in \cite{ghirardo18}, where purely reactive coupling between the cans was assumed, for the case of no flame response.}

\begin{figure}[!t]
\begin{psfrags}
\psfrag{a}{$1.1$}
\psfrag{b}{$1$}
\psfrag{c}{$0.9$}
\psfrag{d}{$-0.2$}
\psfrag{e}{$-0.1$}
\psfrag{f}{$0$}
\psfrag{k}{\hspace{0.2cm}$\nu/\omega_k$}
\psfrag{o}{$\omega/\omega_k$}
\psfrag{p}{$1000$}
\psfrag{q}{$750$}
\psfrag{r}{$500$}
\psfrag{s}{$1.5$}
\psfrag{t}{$1.15$}
\psfrag{u}{$0.8$}
\psfrag{v}{$0.36$ }
\psfrag{w}{$0.21$}
\psfrag{x}{$0.06$ }
\psfrag{A}{\textbf{(a)}}
\psfrag{B}{\textbf{(b)}}
\psfrag{C}{\textbf{(c)}}
\psfrag{D}{$c$ [m/s]}
\psfrag{E}{$L$ [m]}
\psfrag{F}{$B$ [m]}
\psfrag{G}{$b$}
\psfrag{X}{$-0.2$}
\psfrag{Y}{$0.25$}
\psfrag{Z}{$0.15$}
\psfrag{W}{$0.05$}
\psfrag{I}{\textbf{(d)}}
\psfrag{J}{$b=1$}
\psfrag{K}{$0$}
\psfrag{H}{$A$ [m$^2$]}
\includegraphics[width=0.87\textwidth,center]{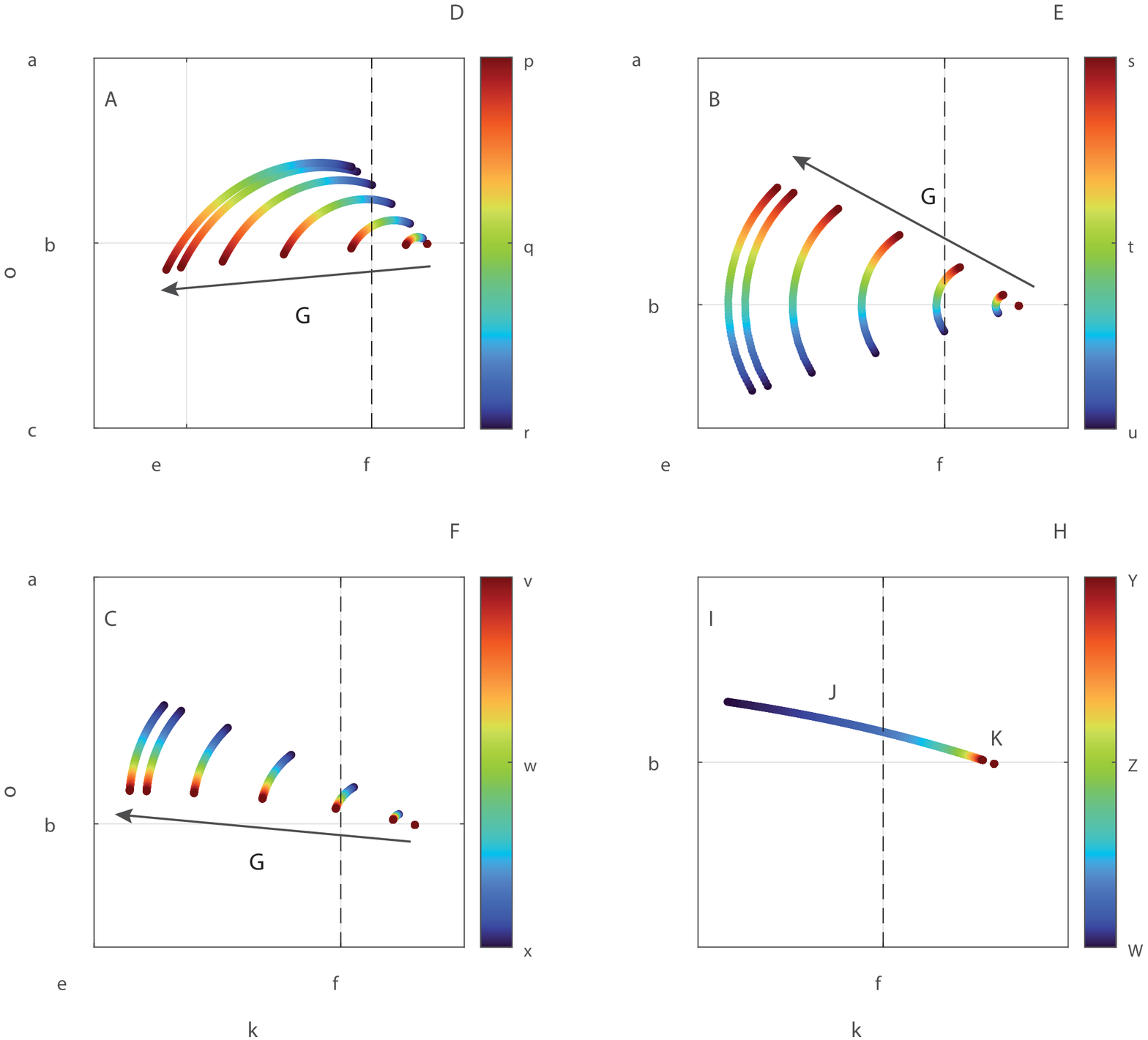}
\end{psfrags}
\caption{\ei{Parameter study in the root locus plane for the first set of parameters in Table \ref{Table 1} and vanishing can spacing $h/W$. The dashed black line marks the stability border $\nu=0$. \ei{The arrows indicate the direction of increasing Bloch wavenumber $b$.} The frequency spectrum $(\omega,\nu)$ is plotted as a function of \textbf{(a)} the ambient speed of sound $c$, \textbf{(b)} the can length $L$, \textbf{(c)} the aperture height $B$ and \textbf{(d)} the cross-section area of the cans $A$. In Fig. \textbf{(a)} and \textbf{(b)}, it was assumed that the eigenfrequency $\omega_k$ varies proportional to $c$ and $1/L$, respectively, starting from the parameter values in Table \ref{Table 1}. In \textbf{(d)}, all higher-order Bloch mode spectra (not shown) follow the same curve as the shown mode with $b=1$.}}\label{Figure 7}
\end{figure}

\zw{The parameter study in the root locus plane is continued in Fig. \ref{Figure 7}, where, for the first set of parameters in Table \ref{Table 1}, the frequency spectrum $(\omega,\nu)$ is plotted as a function of \textbf{(a)} the ambient speed of sound $c$, \textbf{(b)} the can length $L$, \textbf{(c)} the aperture height $B$ and \textbf{(d)} the cross-section area of the cans $A$. \ei{In Fig. \ref{Figure 7}\textbf{(a)} and \textbf{(b)}, it was assumed that $\omega_k$ varies proportional to $c$ and $1/L$, respectively, starting from the parameter values in Table \ref{Table 1}.} In Fig. \ref{Figure 7}\textbf{(d)}, all higher-order Bloch mode spectra (not shown) follow the same curve as the shown mode with $b=1$, but for the same range of values of $A$, they extend farther into the left half-space and end at a lesser growth rate $\nu/\omega_k$.}

\begin{figure}[!t]
\begin{psfrags}
\psfrag{a}{$\nu/\omega_k$}
\psfrag{b}{$\omega/\omega_k$}
\psfrag{c}{$\nu$}
\psfrag{d}{$\omega$}
\psfrag{e}{$\mathrm{St}_k$}
\psfrag{f}{$U$ [m/s]}
\psfrag{g}{$W$ [mm]}
\psfrag{h}{$\nu_0/\omega_k$ [-]}
\psfrag{k}{$1.2$}
\psfrag{l}{$1.1$}
\psfrag{m}{$1$}
\psfrag{n}{$0.9$}
\psfrag{o}{$-0.2$}
\psfrag{p}{$-0.1$}
\psfrag{q}{$0$}
\psfrag{r}{$0.1$}
\psfrag{u}{$4$}
\psfrag{t}{$3$}
\psfrag{s}{$2$}
\psfrag{m}{$1$}
\psfrag{P}{$b$}
\psfrag{v}{$0.8$}
\psfrag{y}{$60$}
\psfrag{x}{$40$}
\psfrag{w}{$20$}
\psfrag{z}{$0.6$}
\psfrag{A}{$0.4$}
\psfrag{C}{$-0.4$}
\psfrag{B}{$-0.6$}
\psfrag{D}{$6$}
\psfrag{E}{$4$}
\psfrag{F}{$2$}
\psfrag{G}{$850$}
\psfrag{H}{$800$}
\psfrag{I}{$750$}
\psfrag{J}{$700$}
\psfrag{K}{$-400$}
\psfrag{L}{$-200$}
\psfrag{N}{$-0.3$}
\psfrag{O}{$0.7$}
\psfrag{F}{$2$}
\psfrag{Z}{$0.3$}
\psfrag{Q}{(a)}
\psfrag{R}{(b)}
\psfrag{S}{(c)}
\psfrag{T}{(d)}
\psfrag{U}{(e)}
\psfrag{V}{(f)}
\psfrag{1}{\hspace{0.2cm}\textcolor{krblue}{$h/W=0$}}
\psfrag{2}{\hspace{0cm}\textcolor{krgreen}{$h/W=0.02$}}
\psfrag{3}{\hspace{0cm}\textcolor{krpurple}{$h/W=0.06$}}
\psfrag{4}{\hspace{0.1cm}\textcolor{black}{$h/W=0.1$}}
\psfrag{5}{\hspace{0.1cm}\textcolor{krcyan}{$h/W=0.2$}}
\psfrag{6}{\hspace{0.1cm}\textcolor{krorange}{$h/W=0.5$}}
\includegraphics[width=0.93\textwidth,center]{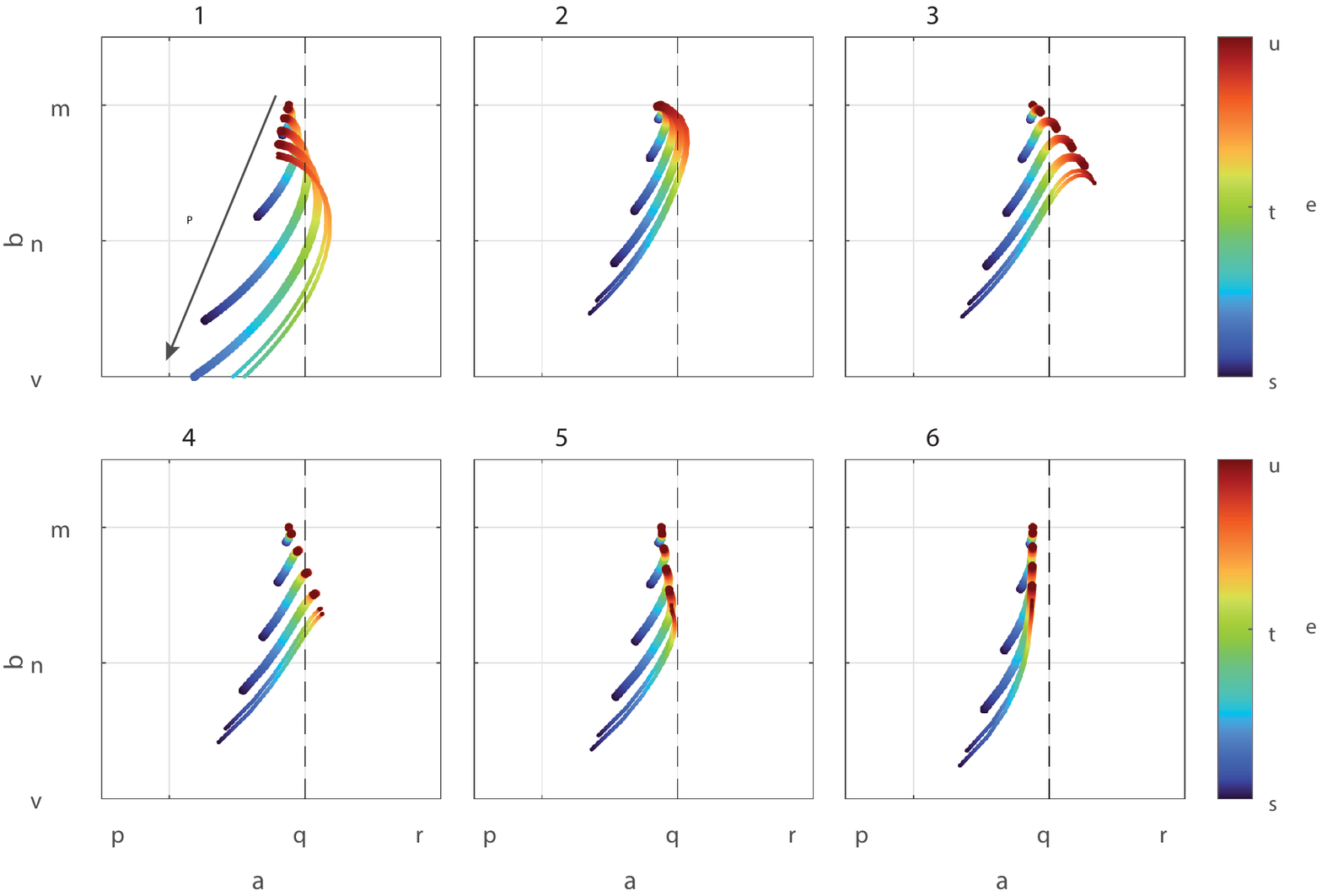}
\end{psfrags}
\caption{\ei{Transition from Fig. \ref{Figure 6}\textbf{(e)} to \textbf{(f)} in more detail. Shown is the frequency spectrum $(\omega,\nu)$ as a function of the normalized eigenfrequency $\mathrm{St}_k=\omega_k W / 2U$ for different values of the can spacing $h/W\in\{0,0.02,0.06,0.1,0.2,0.5\}$. The dashed black line marks the stability border $\nu=0$. The arrow indicates the direction of increasing Bloch wavenumber $b$. The colors above the insets correspond to those of the Rayleigh conductivity curves in Fig. \ref{Figure 3}.}}\label{Figure 8}
\end{figure}

\ei{The influence of the can spacing $h/W$ on the frequency spectrum is investigated in Fig. \ref{Figure 8}, which shows the transition from Fig. \ref{Figure 6}\textbf{(e)} to \textbf{(f)} in more detail. Shown is the spectrum $(\omega,\nu)$ as a function of the normalized eigenfrequency $\mathrm{St}_k=\omega_k W / 2U$ for different values of $h/W\in\{0,0.02,0.06,0.1,0.2,0.5\}$. The colors above the insets correspond to those of the Rayleigh conductivity curves in Fig. \ref{Figure 3}. }

\subsection{Discussion \label{Subsection discussion}}
\zw{We now discuss and give a physical interpretation of the results presented in $\S$\ref{Subsection parameter study}.}

\ei{For nominally unstable conditions, our model predicts that the aeroacoustic coupling between the cans may effectively damp certain Bloch modes over a range of $\omega_k$ around $\omega_k W/U_\mathrm{tot}\approx 0.4$. Under different conditions, the coupling can lead to instabilities in nominally stable systems. Increasing the can width to around $h/W\approx 0.5$ fully suppresses the coupling-induced instability. To the authors' knowledge, these coupling-induced phenomena have not been previously described.}

\ei{We have identified several parameters that influence the stability of the Bloch modes: The eigenfrequency $\omega_k$, the bulk velocity of the combustion products $U_\mathrm{tot}$, the aperture width $W$, the base growth rate $\nu_0$, the ambient speed of sound $c$, the can length $L$, the aperture height $B$ and the cross-section area of the cans $A$.} \zw{In the absence of mean flow in the cans, the coupling between the cans is purely reactive and does not affect the linear stability of the system.}

\zw{We give the following physical interpretation of the results of the parameter study. It is observed in Figs. \ref{Figure 5}, \ref{Figure 6}, \ref{Figure 7} and \ref{Figure 8} that Bloch modes with higher Bloch wavenumber $b$ are more strongly affected by the coupling. To explain this, note that the larger the phase difference between neighboring cans, the larger the apparent acoustic pressure difference at the coupling aperture at a given time. This pressure difference difference drives, through Eq. \eqref{Unsteady Bernoulli equation, not appendix}, the acoustic-hydrodynamic interaction, which may, depending on the value of the Strouhal number, act amplifying or dissipative on the sound field in the control volume. Therefore, the larger the Bloch wavenumber, the stronger the influence of the acoustic coupling between the cans on the linear stability of the Bloch modes.}

\begin{figure}[!t]
\begin{psfrags}
\psfrag{a}{\hspace{0.1cm}$\hat{\eta}_{j}/\hat{\eta}_{j,\mathrm{max}}$}
\psfrag{b}{$\xi$}
\psfrag{f}{$1$}
\psfrag{g}{$0$}
\psfrag{h}{$-1$}
\psfrag{i}{$1$}
\psfrag{j}{$-1$}
\psfrag{K}{$\hspace{3.7cm}$ \begin{tabular}{@{}c@{}}
Bloch mode with $b=5$ \\
Norm. acoustic pressure
\end{tabular}}
\psfrag{L}{$\hspace{3.98cm}$Bloch mode with $b=2$}
\psfrag{N}{$\hspace{2.7cm}$Norm. vortex sheet displacement $\mathrm{Re}\big[\zeta'(t)\big]$}
\psfrag{P}{$\hspace{3.4cm}$Norm. vortex sheet displacement $\zeta'(t)$}
\psfrag{k}{1}
\psfrag{l}{2}
\psfrag{m}{$3$}
\psfrag{n}{$4$}
\psfrag{o}{$5$}
\psfrag{p}{$6$}
\psfrag{q}{$7$}
\psfrag{r}{$8$}
\psfrag{s}{$9$}
\psfrag{t}{$10$}
\psfrag{u}{$11$}
\psfrag{v}{$12$}
\psfrag{A}{\textbf{(a)}}
\psfrag{B}{\textbf{(b)}}
\psfrag{C}{$\pi$}
\psfrag{D}{$\dfrac{3\pi}{2}$}
\psfrag{1}{\hspace{0.3cm}\ei{Can number $j$}}
\psfrag{X}{$\textcolor{red}{\zeta'(t)}$}
\psfrag{Y}{$\textcolor{red}{\zeta'(t)}$}
\psfrag{M}{\hspace{0.3cm}$7$}
\psfrag{Q}{\textbf{(a)}}
\psfrag{R}{\textbf{(b)}}
\psfrag{S}{\textbf{(c)}}
\psfrag{T}{\textbf{(d)}}
\includegraphics[width=0.93\textwidth,center]{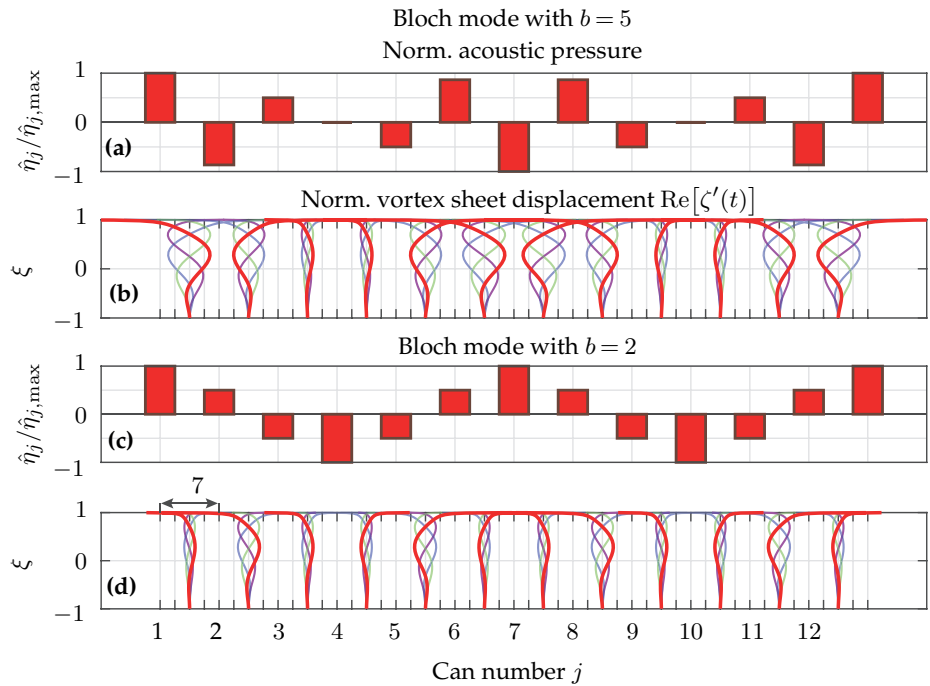}
\end{psfrags}
\caption{\zw{Visualization of the acoustic-hydrodynamic interaction corresponding to a coupling-induced instability for the second set of parameters in Table \ref{Table 1}.} \ei{Vanishing can spacing $h/W$ was assumed.} \zw{Shown in \textbf{(a)} and \textbf{(b)} are the normalized acoustic pressure distribution at a given time instance and the real part of the normalized vortex sheet displacement $\mathrm{Re}\big[\zeta' e^{(-\mathrm{i}\omega  + \nu) t}\big]$ at 4 equally spaced points in time during an acoustic cycle with period $T=2\pi/\omega$, respectively, for a Bloch mode with $b=5$. \gc{For visualization purposes, $\zeta'$ is scaled with the normalized pressure difference between the cans.} Shown in \textbf{(c)} and \textbf{(d)} is the same for a Bloch mode with $b=2$. The normalized frequency spectrum $(\omega/\omega_k,\nu/\omega_k)$ is $(0.899,1.53\times 10^{-2})$ for $b=5$ and $(0.978,-6.19\times 10^{-3})$ for $b=2$. Note that the periods $T$ are different for the two cases shown in \textbf{(b)} and \textbf{(d)}, respectively, and that the Bloch mode with $b=2$ oscillates at a $9\%$ higher frequency than the one with $b=5$.}}\label{Figure 9}
\end{figure}

\ei{This interpretation is exemplified in Fig. \ref{Figure 9}, which visualizes the acoustic-hydrodynamic interaction corresponding to a coupling-induced instability for the second set of parameters in Table \ref{Table 1}. For simplicity, vanishing can spacing $h/W$ was assumed.} \zw{Shown in Fig. \ref{Figure 9}\textbf{(a)} and \textbf{(b)} are the normalized acoustic pressure distribution at a given time instant and the real part of the normalized vortex sheet displacement $\mathrm{Re}\big[\zeta' e^{(-\mathrm{i}\omega  + \nu) t}\big]$ at 4 equally spaced points in time during an acoustic cycle with period $T=2\pi/\omega$, respectively, for the Bloch mode with $b=5$. \gc{For visualization purposes, $\zeta'$ is scaled with the normalized pressure difference between the cans.} The insets in Fig. \ref{Figure 9}\textbf{(c)} and \textbf{(d)} show the same for the Bloch mode with $b=2$. In these cases, the normalized frequency spectrum $(\omega/\omega_k,\nu/\omega_k)$ is $(0.899,1.53\times 10^{-2})$ for $b=5$ and $(0.978,-6.19\times 10^{-3})$ for $b=2$. In this example, the higher apparent pressure differences across the coupling interfaces lead to an instability of a higher-order Bloch mode, while the lower-order Bloch mode remains stable. Note that the periods $T$ are different for the cases shown in Fig. \ref{Figure 9} \textbf{(b)} and \textbf{(d)}, respectively, and that the Bloch mode with $b=2$ oscillates at a $9\%$ higher frequency than the one with $b=5$.}

\zw{The large displacements of the vortex sheet at the downstream edge of the aperture shown in Fig. \ref{Figure 9} are characteristic of Howe's theory (see p. 437 in \cite{howe_1998}). This typical spatial behavior of the vortex sheet displacement, which is enabled by the Kutta condition \eqref{Kutta condition}, is a simplified representation of the violent motions and acoustic energy production that occurs when pockets of coherent vorticity shed from the upstream edge make contact with the downstream edge \cite{howe_1981}. In reality, shedding of discrete vortices can take place where the shear layer rolls up before the turbine inlet. Vortex sheet roll-up has been revisted recently by \cite{Devoria2018299}. A numerical study on acoustic sound production by grazing turbulent flow over a T-junction aperture including examples of discretely shed vortices is presented in \cite{Bauerheim20}.}

\section{Conclusions} \label{Section 6: Conclusions}
\zw{We have derived a coupled oscillator model of a thermoacoustic instability in an idealized can-annular combustor. By combining the unimodal projection of the Helmholtz equation for the can acoustics, a detailed fluid-dynamical model for the can-to-can communication and a Bloch wave ansatz, we derived a single equation for the frequency spectrum. We performed a parameter study and identified two special conditions: one where the aeroacoustic coupling acts dissipative on the nominally unstable thermoacoustic system and one where amplifying coupling leads to an instability in a nominally stable system. We identified several model parameters which influence the system stability, including the bulk velocity of the combustion products $U_\mathrm{tot}$, whose effect on the system stability has not been considered in previous studies. We gave a physical interpretation of our results, arguing that higher-order Bloch modes more strongly drive the acoustic-hydrodynamic interaction between the cans due to higher apparent pressure differences at the coupling interfaces. This leads to a stronger influence of the coupling on these higher-order modes. We believe the present analysis, which highlights the effect of the fluid motion in the apertures between the cans on the thermoacoustic instability, can further the rational development of mitigation measures against instabilities in real-world gas turbines.}
\vskip6pt

\enlargethispage{20pt}


\dataccess{The datasets used for generating the plots and
results in the present study can be directly obtained from the
numerical simulation of the related mathematical equations in
the manuscript.}

\aucontribute{T. P. carried out the formal analysis and the investigation, performed the simulations and the model validation, wrote the original draft and revised the manuscript. N. N. conceived and supervised the study, helped carry out the formal analysis and the investigation, critically reviewed and edited the original draft. All authors gave final approval for publication and agree to be held accountable for the work performed therein.}

\competing{The authors declare that they have no competing interests.}

\funding{This project is funded by the Swiss National Science Foundation under Grant agreement 184617. }




\vskip2pc

\bibliographystyle{RS} 

\bibliography{sample} 

\end{document}